\documentclass[onecolumn,preprintnumbers,amsmath,amssymb]{revtex4-1}
\usepackage{graphicx}
\usepackage{dcolumn}
\usepackage{bm}
\usepackage{amsmath}
\usepackage[colorlinks]{hyperref}
\newcommand{\bef}{\begin{figure}}
\newcommand{\eef}{\end{figure}}

\newcommand{\be}{\begin{equation}}
\newcommand{\ee}{\end{equation}}
\newcommand{\bea}{\begin{eqnarray}}
\newcommand{\eea}{\end{eqnarray}}
\begin{document}

\title{The first moment of azimuthal anisotropy in nuclear collisions from AGS to LHC energies}

\author{Subhash Singha, Prashanth Shanmuganathan and Declan Keane}
\affiliation{Department of Physics, Kent State University, Ohio 44242, USA}

\date{\today}

\begin{abstract}
We review topics related to the first moment of azimuthal anisotropy
($v_1$), commonly known as directed flow, focusing on both charged
particles and identified particles from heavy-ion collisions.  Beam
energies from the highest available, at the CERN LHC, down to
projectile kinetic energies per nucleon of a few GeV per nucleon, as
studied in experiments at the Brookhaven AGS, fall within our scope. We focus on experimental measurements and on theoretical work where direct comparisons with experiment have been emphasized. The physics addressed or potentially addressed by this review topic includes the study of Quark Gluon Plasma, and more generally, investigation of the Quantum Chromodynamics phase diagram and the equation of state describing the accessible phases. 
\end{abstract}
\maketitle

\section{Introduction}
\label{s-Intro}
The purpose of relativistic nuclear collision experiments is the creation and 
study of nuclear matter at high energy densities. Experiments have 
established a new form of strongly-interacting matter, called Quark Gluon
Plasma (QGP)~\cite{WhiteP1, WhiteP2, WhiteP3, WhiteP4, QGP_review2006, 
QGP_review2010, RecentQGP}. Collective motion of the particles emitted from 
such collisions is of special interest because it is sensitive to the equation of 
state in the early stages of the reaction~\cite{h_ritter_review, earlytime, 
Norbert_review}. Directed flow was the first type of collective motion identified 
among the fragments from nuclear collisions \cite{Miklos1, Miklos2, PlasticBall}, 
and in current analyses, is characterized by the first harmonic coefficient in the 
Fourier expansion of the azimuthal distribution of the emitted particles with 
respect to each event's reaction plane azimuth ($\Psi$)~\cite{flow_method1, 
flow_method2, flow_method3}:

\begin{equation}
v_1 =\langle\cos(\phi-\Psi)\rangle,
\label{eq1}
\end{equation}
where $\phi$ is the azimuth of a charged particle, or more often, the 
azimuth of a particular particle species, and the angle brackets denote 
averaging over all such particles in all events. In some experimental analyses, 
$v_1$ is evaluated directly from Eq.~(\ref{eq1}), then a correction is applied for 
reaction plane resolution \cite{flow_method2}, whereas in a typical modern   
analysis method, the directed flow correlation is extracted using cumulants 
\cite{flow_method3}.  In general, $v_1$ is of 
interest when plotted as a function of rapidity, $y$, or sometimes 
pseudorapidity $\eta = -\ln(\tan\theta/2)$, where $\theta$ is the polar 
angle of the particle. The dependence of $v_1$ on collision centrality 
and on transverse momentum, $p_T$, can offer additional insights.  
 
Until relatively recently \cite{v1even1, v1even2}, the rapidity-even component 
$v_1^{\rm even}(y) = v_1^{\rm even}(-y)$ was always assumed to be zero or 
negligible in mass-symmetric collisions. In fact, fluctuations within the 
initial-state colliding nuclei, unrelated to the reaction plane, can generate 
a significant $v_1^{\rm even}$ signal \cite{v1even1, v1even2}. This fluctuation 
effect falls beyond the scope of the present review, which focuses on fluid-like 
directed flow, $v_1^{\rm odd}(y) = - v_1^{\rm odd}(-y)$, as per Eq.~(\ref{eq1}), 
and from here on, $v_1$ for mass-symmetric collisions implicitly signifies 
$v_1^{\rm odd}$.  

During the first decade of the study of $v_1$ in nuclear collisions, it was 
more commonly called sideward flow. It refers to a sideward collective 
motion of the emitted particles, and is a repulsive collective deflection  
in the reaction plane. By convention, the positive direction of $v_1$ is 
taken to be the direction of `bounce-off' of projectile spectators in a 
fixed target experiment \cite{h_ritter_review, Norbert_review}. Models 
imply that directed flow, especially the component closest to beam 
rapidities, is initiated during the passage time of the two colliding 
nuclei; the typical time-scale for this is $2R/\gamma$ \cite{earlytime, 
Norbert_review}, where $R$ and $\gamma$ are the nuclear radius and 
Lorentz factor, respectively. This is even earlier than the still-early time 
when elliptic flow, $v_2$, is mostly imparted. Thus $v_1$ can probe the 
very early stages of the collision \cite{jam_attractive, phsd}, 
when the deconfined state of quarks and 
gluons is expected to dominate the collision dynamics~\cite{earlytime, 
Norbert_review}. Both hydrodynamic~\cite{hydro_heinz, stocker_npa_750} 
and transport model \cite{urqmd1, urqmd2} calculations indicate that the 
directed flow of charged particles, especially baryons at midrapidity, is 
sensitive to the equation of state and can be used to explore the QCD phase 
diagram. 

The theoretical work leading to the prediction of collective flow in nuclear 
collisions evolved gradually. In the mid-1950s, Belenkij and Landau 
\cite{Landau1956} were the first to consider a hydrodynamic description of 
nuclear collisions. During the 1970s, as the Bevatron at Lawrence Berkeley 
National Lab was converted for use as the first accelerator of relativistic 
nuclear beams, the idea of hydrodynamic shock compression of nuclear 
matter emerged~\cite{Chapline1973, Frankfurt1974, Frankfurt1976}, 
and these developments in turn led to increasingly realistic predictions 
\cite{Frankfurt1980, Miklos1, Miklos2} that paved the way for the first 
unambiguous measurement of directed flow at the Bevalac in the 
mid-1980s \cite{PlasticBall}. A frequent focus of theory papers during the 
subsequent years was the effort to use directed flow measurements to infer 
the incompressibility of the nuclear equation of state in the hadron gas 
phase and to infer properties of the relevant momentum-dependent potential 
\cite{h_ritter_review, Norbert_review}. The observed directed flow at AGS 
energies \cite{E877_npa_590, E877_prc_56, E877_prc_59, E877_plb_485, 
E895_prl_84, E895_prl_85, E895_prl_86, E895_npa_698} and below 
is close to a linear function of rapidity throughout the populated region, 
and the slope $dv_1/dy$ can adequately quantify the strength of the signal.  
At SPS energies and above~\cite{na49_prl_80, na49_prc_68, star_prc_72, 
phobos_prl_97, star_prc_73, star_prc_85, star_prl_101, star_prl_108, 
star_prl_112}, a more complex structure is observed in $v_1(y)$, with the 
slope $dv_1/dy$ in the midrapidity region being different from the slope in 
the regions closer to beam rapidities.  

Various models also exhibit this kind of behavior. 
At these energies, both hydrodynamic and nuclear transport calculations predict 
a negative sign for charged particle $dv_1/dy$ near midrapidity, where pions are 
the dominant particle species. This negative $dv_1/dy$ near midrapidity has been 
given various names in the literature: ``third flow component'' \cite{Csernai}, 
``anti-flow'' \cite{Brachmann}, or ``wiggle'' \cite{Norbert_review, 
snellings_prl_84, stocker_npa_750}. This phenomenon has been discussed as a 
possible QGP signature, and a negative $dv_1/dy$ for baryons has been argued 
\cite{stocker_npa_750} to be particularly significant. However, some aspects 
of anti-flow can be explained in a model with only hadronic physics 
\cite{snellings_prl_84, Aihong-v1Model} by assuming either incomplete baryon 
stopping with a positive space-momentum correlation~\cite{snellings_prl_84}, 
or full stopping with a tilted source~\cite{bozek_prc_81}. 

A three-fluid hydrodynamic model~\cite{stocker_npa_750} predicts a monotonic 
trend in net-baryon directed flow versus beam energy in the case of a purely 
hadronic equation of state, whereas a prominent minimum at AGS energies, 
dubbed the ``softest point collapse'', is predicted when the equation of state 
incorporates a first-order phase transition between hadronic and quark-gluonic 
matter. Recent measurements of both proton and net-proton directed flow at 
RHIC~\cite{star_prl_112} indeed indicate non-monotonic directed flow as a 
function of beam energy, with the minimum lying between 11.5 and 19.6 GeV in 
$\sqrt{s_{NN}}$. However, more recent hydrodynamic and nuclear transport 
calculations which incorporate significant theoretical improvements (see Section 
\ref{s-Models}) do not reproduce the notable qualitative features of the data, 
and therefore cast doubt on any overall conclusion about the inferred properties 
of the QCD phase diagram. Directed flow has also been measured at the
LHC \cite{alice_prl_111}. A negative slope of $v_1(\eta)$ is observed for charged 
particles, but its magnitude is much smaller than at RHIC, which is thought to be 
a consequence of the smaller tilt of the participant zone at the LHC.

In this article, we review a representative set of directed flow results spanning 
AGS to LHC energies. In Sections \ref{s-Charged} and \ref{s-CuAu}, we discuss 
measurements of $v_1$ for charged particles in mass-symmetric and 
mass-asymmetric collisions, respectively. In Section \ref{s-PID}, we cover 
measurements of $v_1$ for various identified particle species. Section \ref{s-Models} 
reviews some recent model calculations which lend themselves to direct comparisons 
with directed flow data. Section \ref{s-End} presents a summary and future outlook.


\section{Differential measurements of charged particle directed flow }
\label{s-Charged}
In this section, we review measurements of $v_1$ for all charged particles in 
cases where individual species were not identified. Studies of the dependence 
on transverse momentum $p_T$, pseudorapidity $\eta$, beam energy $\sqrt{s_{NN}}$, 
system size, and centrality are included.

\subsection{Dependence of $v_1$ on transverse momentum}

\begin{figure}
\begin{center}
\includegraphics[scale=0.4]{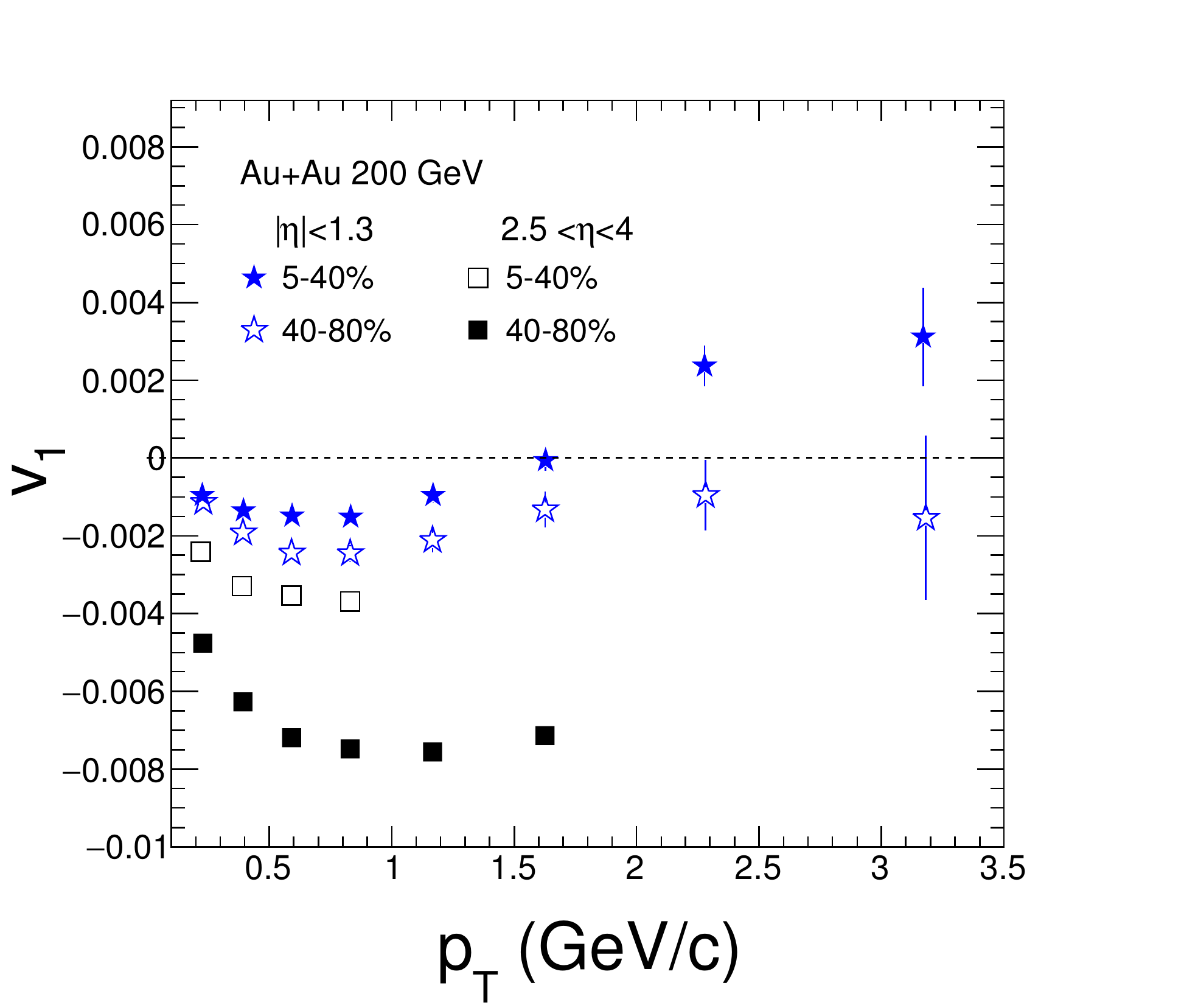}
\includegraphics[scale=0.4]{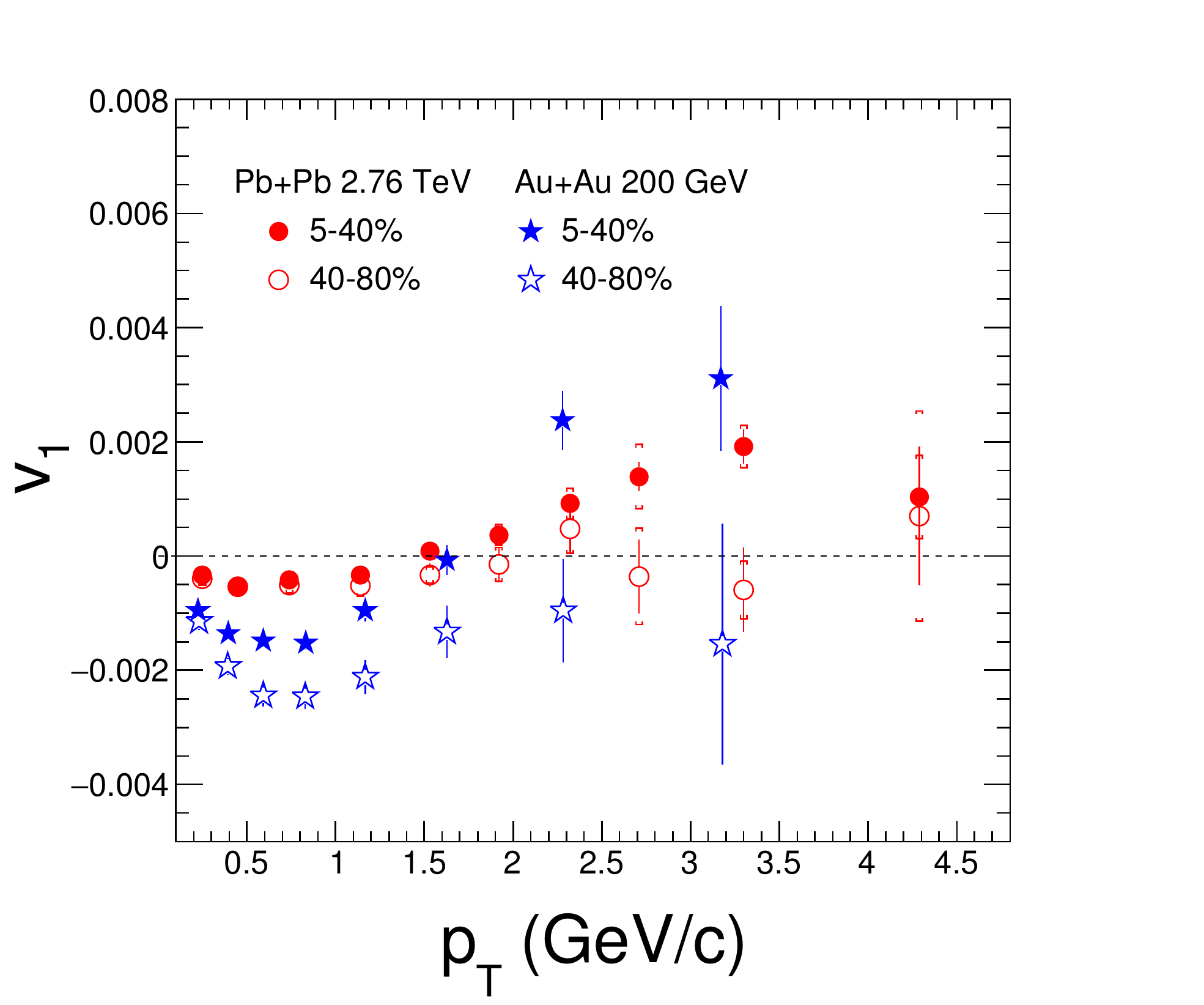}
\caption{(Color online) Left panel: Charged particle $v_1$ as a function of transverse 
momentum in 200 GeV Au+Au collisions at RHIC, for two centralities and two 
pseudorapidity windows ($|\eta| < 1.3$ and $2.5 < \eta < 1.3$)~\cite{star_prl_101}. 
Right panel: Comparison of RHIC results with midrapidity measurements in 
2.76 TeV Pb+Pb collisions at the LHC~\cite{alice_prl_111}.}
\label{v1pt}
\end{center}
\end{figure} 

The $p_T$-dependence of $v_1$ for charged particles has been studied by the STAR 
experiment at RHIC~\cite{star_prc_73, star_prl_101}. The left panel of Fig.~\ref{v1pt} 
presents directed flow results for Au+Au collisions in two centrality intervals: 
5-40\% and 40-80\%, and in two regions of pseudorapidity: $|\eta| < 1.3$ and 
$2.5 < |\eta| < 1.3$.  In this case, because of the odd-functional property of 
$v_1(\eta)$, the backward pseudorapidity region by convention has its 
sign of $v_1$ reversed before summing over the indicated gate in pseudorapidity. 
These measurements represent the first instance of using only spectators to 
determine the estimated azimuth of the reaction plane~\cite{ZDC-SMD_STARnote, 
Gang_PhD, ZDC-SMD_hardware}.   

The measured $v_1$ in the mid-pseudorapidity interval $|\eta| < 1.3$ crosses zero 
in the $p_T$ region of 1 to 2 GeV/$c$. Extrapolations raise the possibility that a 
qualitatively similar zero crossing by forward-pseudorapidity $v_1$ occurs at higher 
$p_T$, but that region does not fall within the acceptance of the STAR detector. 

The charged particle $v_1(p_T)$ measurements by ALICE~\cite{alice_prl_111} in Pb+Pb 
collisions at 2.76 TeV are compared in the right panel of Fig.~\ref{v1pt} with the 
corresponding data from 200 GeV Au+Au collisions at RHIC. The measurements at both 
LHC and RHIC shows a similar trend, including a sign change around $p_T \sim 1.5$ 
GeV/$c$ in central collisions and negative values at all $p_T$ for peripheral
collisions. 

There is an interest in the observation of zero crossing behavior in $v_1(p_T)$ as 
it can be used to constrain hydrodynamic model calculations \cite{heinz_jpg_30}.
It has been pointed out that this sign change is an artifact of combining 
all species of charged particles together, and can be explained \cite{star_prl_101} 
by the different sign of $v_1$ for pions and baryons, in conjunction with the 
enhanced production of baryons at higher $p_T$ \cite{star_plb_655}.  This 
complication is one of the reasons why directed flow of identified particles, as 
reviewed in later sections, can be easier to interpret and offers additional insights.  

\subsection{Dependence of $v_1$ on pseudorapidity}

\begin{figure}
\begin{center}
\includegraphics[scale=0.4]{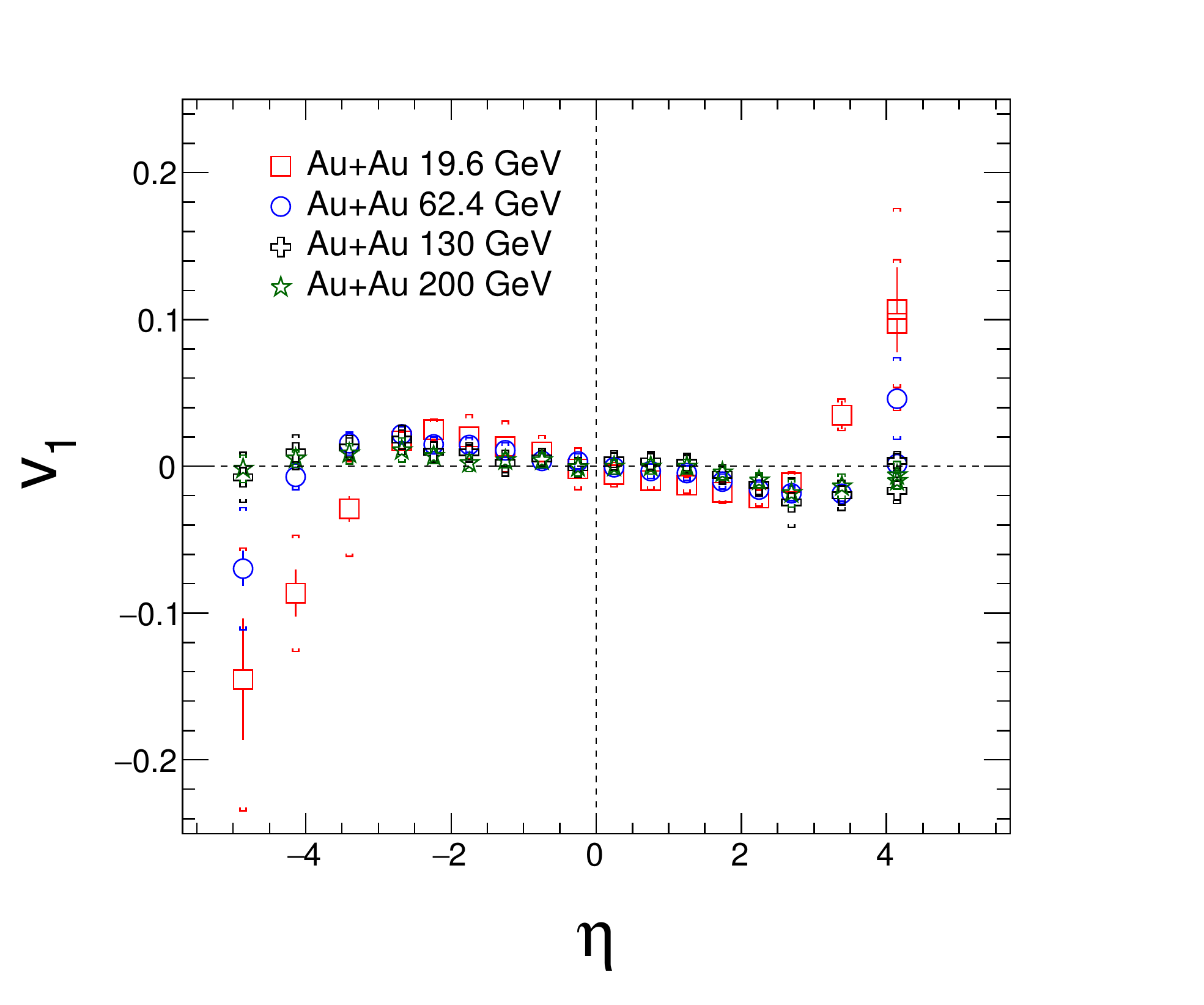}
\includegraphics[scale=0.4]{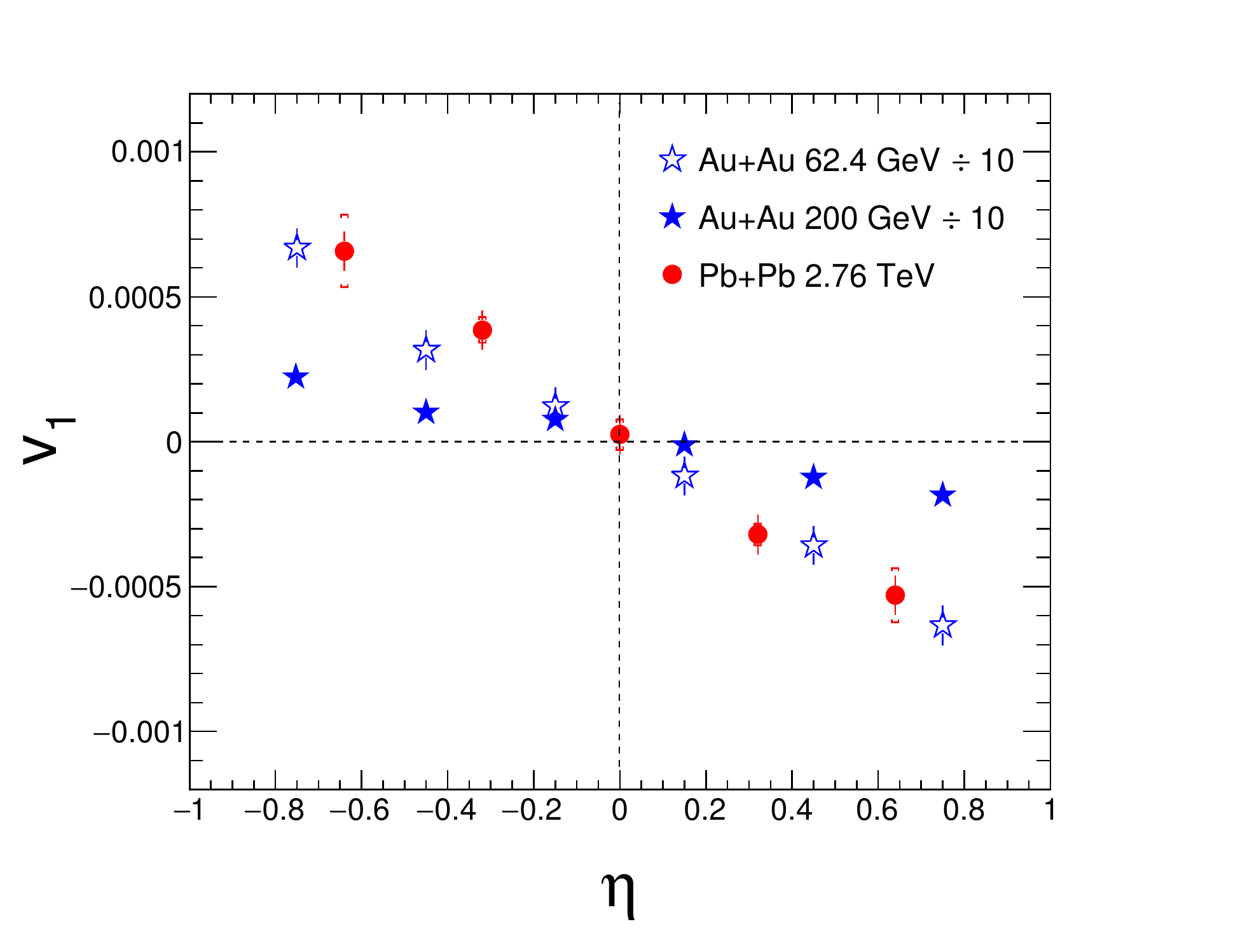}
\caption{(Color online) Left panel: Charged particle $v_1$ as a function of $\eta$ 
  for 0-40\% central Au+Au collisions at $\sqrt{s_{NN}} = 19.6$, 62.4, 130 and 200
  GeV, measured by PHOBOS at RHIC~\cite{phobos_prl_97}. Right panel: Comparison 
  of charged particle $v_1$ vs.\,$\eta$ at RHIC~\cite{star_prl_101} and 
  LHC~\cite{alice_prl_111} energies. Note that all the RHIC data points are 
  divided by 10 in order to be plotted on a common scale with LHC results. }
\label{v1_eta_star_alice}
\end{center}
\end{figure}

The left panel in Fig.~\ref{v1_eta_star_alice} shows charged particle $v_1(\eta)$ 
in Au+Au collisions at 19.6, 62.4, 130 and 200 GeV measured in the PHOBOS 
detector~\cite{phobos_prl_97}. It is evident that charged particle $v_1$ within 
3 to 4 units of $\eta$ on either side of $\eta = 0$ has a sign opposite to that of
the spectators on that side of $\eta = 0$ (the anti-flow phenomenon). 

The right panel of Fig.~\ref{v1_eta_star_alice} shows the $\eta$ dependence of
$v_1$ for charged particles in 2.76 TeV Pb+Pb collisions, as measured by the ALICE 
collaboration~\cite{alice_prl_111} at the LHC. The ALICE results are compared in 
this panel with RHIC measurements from STAR~\cite{star_prc_72, star_prl_101}. The 
$v_1$ slope at the LHC and at the top RHIC energies has the same negative sign, but 
the slope magnitude at the LHC is a factor $\sim 3$ smaller than at top RHIC energy. 
This pattern is consistent with the participant zone at the LHC having a smaller 
tilt, as predicted~\cite{bozek_prc_81}, and does not support a proposed picture at 
LHC energies in which a strong rotation is imparted to the central fireball 
\cite{Bleibel2008, Lazlo2011}.

\subsection{System size and beam energy dependence of $v_1$}

\begin{figure}
\begin{center}
\includegraphics[scale=0.4]{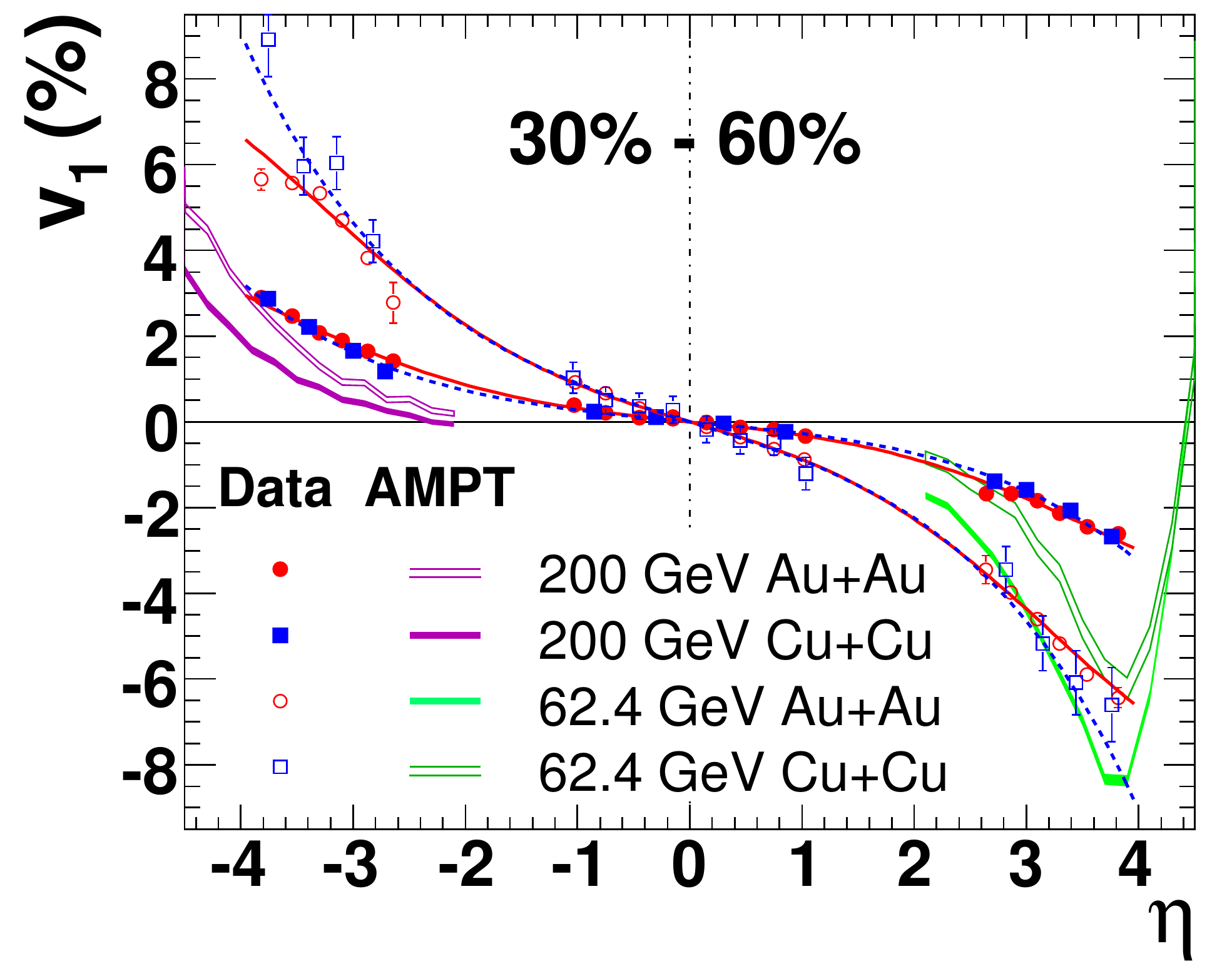}
\includegraphics[scale=0.4]{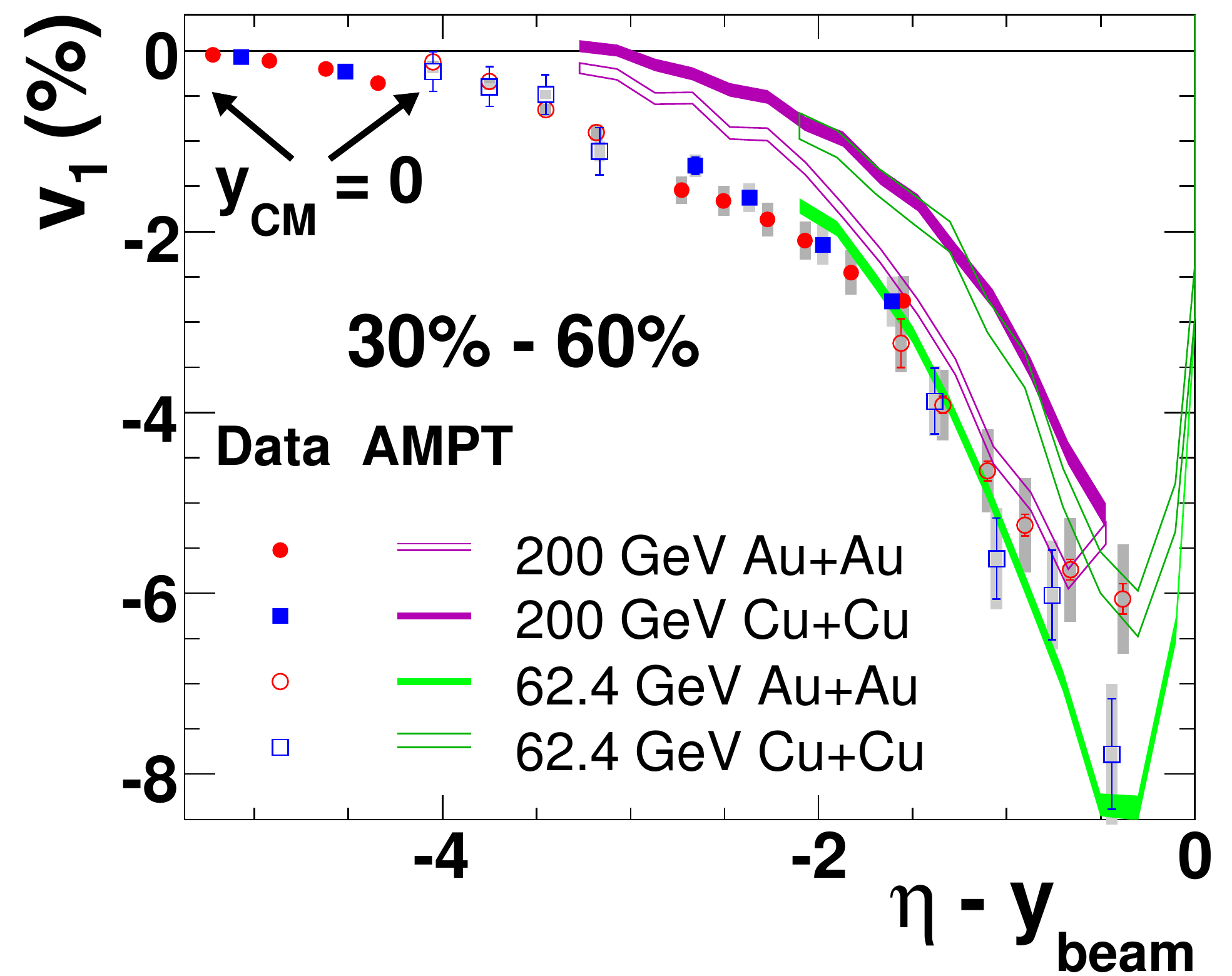}
\caption{(Color online) Left panel: Charged particle $v_1$ as a function of
  $\eta$ for 200 and 62.4 GeV Au+Au and Cu+Cu collisions~\cite{star_prl_101}. 
  Right panel: the same as the left panel, but with the $x$-axis shifted by 
  $y_{\rm beam}$.}
\label{v1_eta_auau_cucu}
\end{center}
\end{figure}

The beam energy and system size dependence of $v_1$ have been studied at
RHIC using data from two colliding species at two beam energies. The left 
panel of Fig.~\ref{v1_eta_auau_cucu} shows charged particle $v_1$ for 
mid-centrality (30-60\%) Au+Au and Cu+Cu collisions at $\sqrt{s_{NN}} = 62.4$
and 200 GeV, measured by STAR~\cite{star_prl_101}. A trend of decreasing 
$v_1(\eta)$ is observed as beam energy increases for both Au+Au and Cu+Cu
collisions. Across the reported pseudorapidity range, $v_1(\eta)$ is 
independent of system size within errors at each beam energy. This is a
remarkable finding, given that the Au+Au system mass is three times that of 
the Cu+Cu system, and given that neither the AMPT~\cite{ampt1,ampt2,ampt3} 
nor the UrQMD~\cite{urqmd1,urqmd2} models exhibit such a scaling behavior.

A different scaling behavior is presented in the right panel of 
Fig.~\ref{v1_eta_auau_cucu}. Here, the data in the left panel are transformed 
into the rest frame of the beam nucleus, i.e., zero on the $x$-axis corresponds 
to $y_{\rm beam}$ for each of the two collision energies involved. Within 
errors, the measurements lie on a universal curve across about three units of 
pseudorapidity. This behavior is known in the heavy-ion literature as limiting 
fragmentation, and had previously been observed in Au+Au collisions as a function 
of beam energy by STAR \cite{star_prc_73} and PHOBOS \cite{phobos_prl_97}.  
The term `limiting fragmentation' was originally employed by Feynman \cite{LimFrag} 
and Benecke {\it et al.} \cite{LimFrag2} to describe the analogous phenomenon of 
the measured $\mu^+/\mu^-$ ratio in cosmic ray showers at sea level being almost 
independent of muon energy.

\subsection{Centrality dependence of $v_1$}

\begin{figure}
\begin{center}
\includegraphics[scale=0.5]{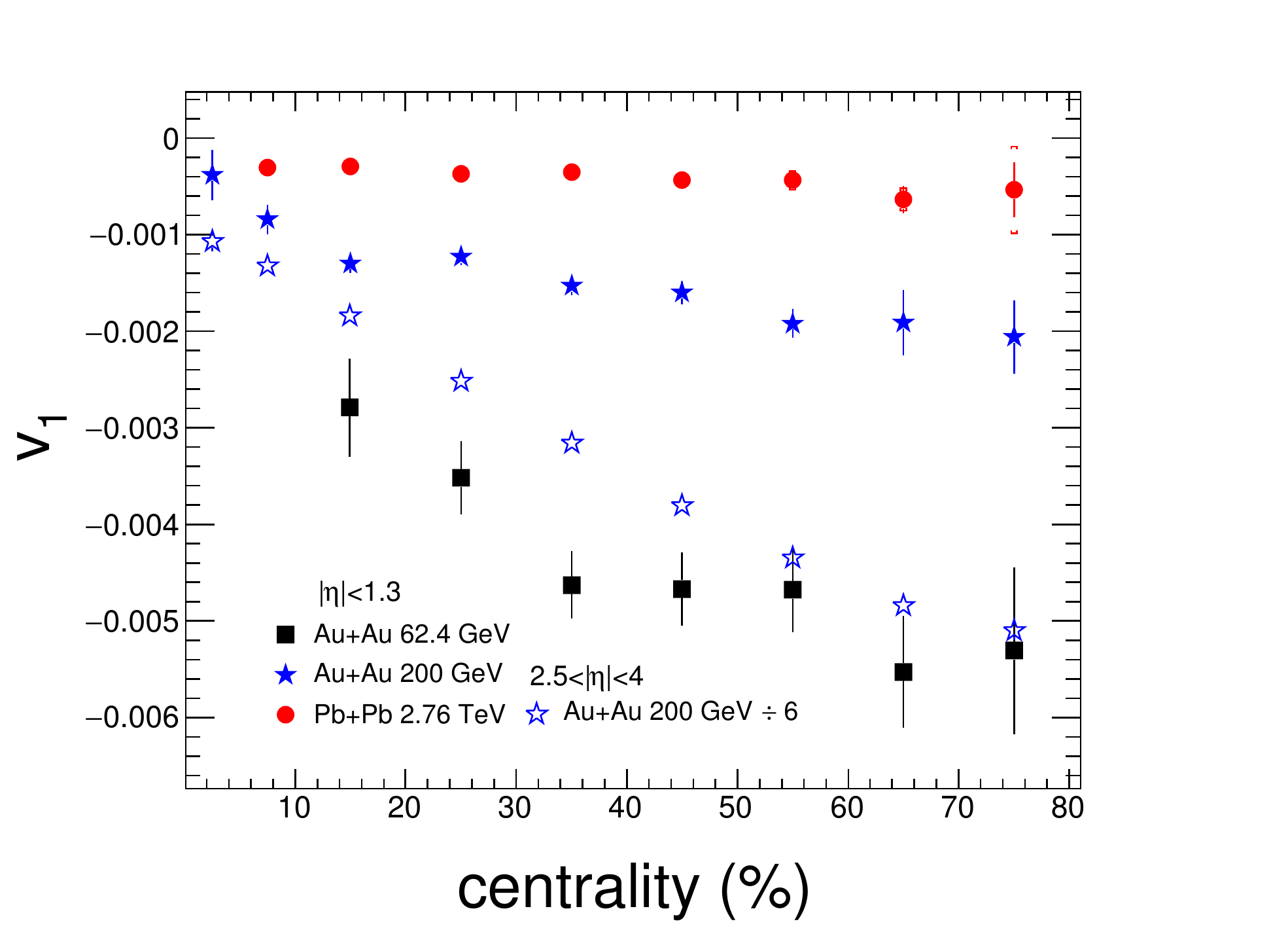}
\caption{(Color online) $v_1$ of charged particles as a function of centrality 
  for a mid-pseudorapidity region ($|\eta| < 1.3$) and a forward pseudorapidity 
  region ($2.5 < |\eta| < 4.0$). Two RHIC datasets, 62.4 and 200 GeV Au+Au 
  \cite{star_prl_101} and one LHC dataset, 2.76 TeV Pb+Pb \cite{alice_prl_111} 
  are shown at mid-pseudorapidity, while only 200 GeV Au+Au \cite{star_prl_101} 
  is shown at forward pseudorapidity. Note that the data points at forward 
  pseudorapidity are divided by 6 in order to be plotted on a common scale with 
  the data at mid-pseudorapidity.  }
\label{v1_centrality_star_alice}
\end{center}
\end{figure} 
Fig.~\ref{v1_centrality_star_alice} shows $v_1$ as a function of collision 
centrality in 62.4 and 200 GeV Au+Au \cite{star_prl_101} from STAR and in 
2.76 TeV Pb+Pb \cite{alice_prl_111} from ALICE, for the mid-pseudorapidity 
region $|\eta| < 1.3$.  Fig.~\ref{v1_centrality_star_alice} also shows $v_1$ 
(divided by 6 in order to fit conveniently on a common scale) as a function 
of collision centrality in 200 GeV Au+Au \cite{star_prl_101} from STAR at the 
forward pseudorapidity region $2.5 < |\eta| < 4.0$. 

Directed flow magnitude at mid-pseudorapidity increases monotonically going 
from central to peripheral collisions at all three beam energies, and there is 
a strong trend for this magnitude to decrease with increasing beam energy.  A 
similar but stronger centrality dependence is observed at forward pseudorapidity,
and the magnitude also increases strongly from mid to forward pseudorapidity. 

It has been pointed out by Caines \cite{Helen2006} that many aspects 
of soft physics (defined as $p_T < 2$ GeV/$c$) in heavy-ion collisions at 
relativistic energies depend only on event multiplicity per unit rapidity; in 
other words, for a fixed value of $dN_{\rm ch}/d\eta$, there is no significant 
dependence on beam energy, or on centrality, or on the mass of the colliding 
system. This type of scaling is called ``entropy-driven'' soft physics. The 
directed flow results for two beam energies and two colliding systems reported 
by STAR in Ref.~\cite{star_prl_101} (see Fig.~\ref{v1_eta_auau_cucu}) 
represented one of the first (and still few) violations of entropy-driven 
multiplicity scaling. In contrast, this scaling is observed to hold, with 
caveats, for homogeneity lengths from femtoscopy \cite{Lisa2005, Helen2006}, 
for elliptic flow per average participant eccentricity \cite{Helen2006, 
PHOBOS_v2}, and for various strangeness yields \cite{Helen2006}. 
More recent data from the LHC also reveal examples of entropy-driven 
multiplicity scaling \cite{Helen2016}.


\section{Directed flow of charged particles in mass-asymmetric collisions }
\label{s-CuAu}

\begin{figure}
\begin{center}
\includegraphics[scale=0.8]{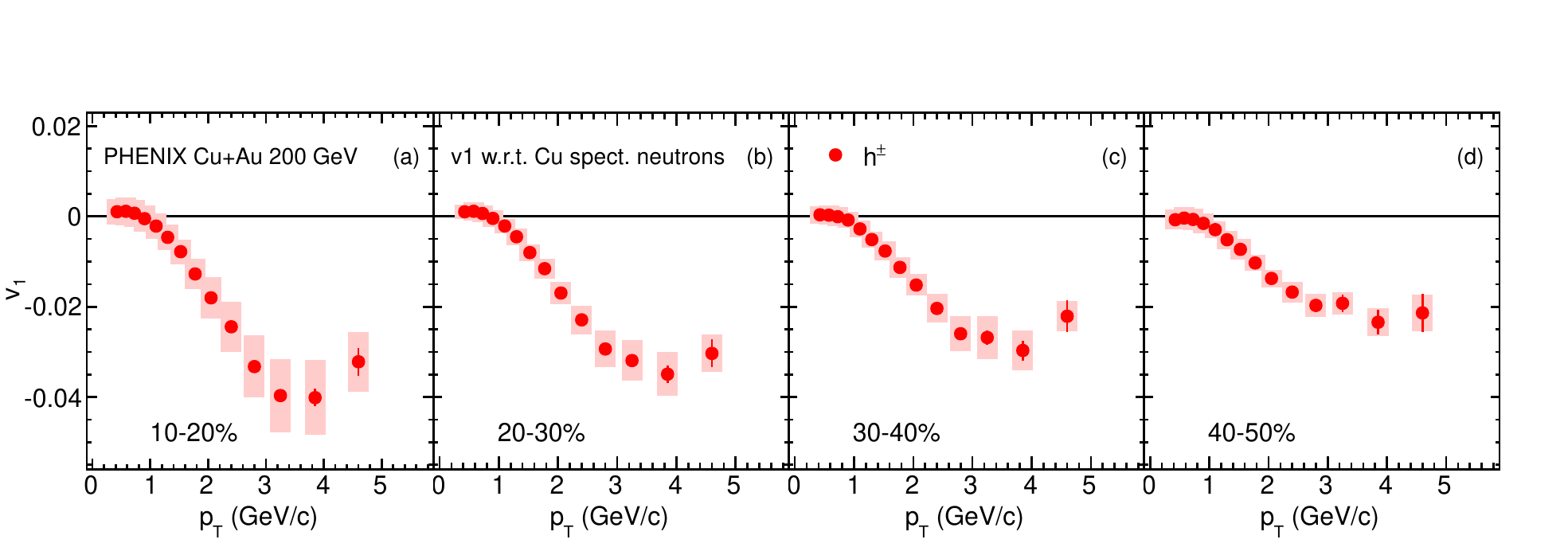}
\caption{(Color online) Midrapidity $v_1(p_T)$ in Cu+Au collisions at 
$\sqrt{s_{NN}} = 200$ GeV in four bins of centrality, as reported by the 
PHENIX collaboration~\cite{phenix_CuAu}. }
\label{phenix_CuAu_pT}
\end{center}
\end{figure} 

In mass-asymmetric collisions like Cu+Au, the well-defined distinction that exists 
in mass-symmetric collisions between the odd $v_1(\eta)$ component (a hydrodynamic 
effect correlated with the reaction plane) and the even $v_1(\eta)$ component 
(an initial-state fluctuation effect unrelated to the reaction plane) no longer 
holds. In a recent paper from the PHENIX collaboration, they report midrapidity 
charged hadron $v_1(p_T)$ in Cu+Au collisions at $\sqrt{s_{NN}} = 200$ GeV for 
centralities of 10-20\%, 20-30\%, 30-40\% and 40-50\%, using spectator neutrons 
from the Au side of the collision to determine the event plane; see Fig.~\ref
{phenix_CuAu_pT}~\cite{phenix_CuAu}. However, they preserve the standard convention 
for the sign of $v_1$ by defining the direction of bounce-off by remnants of the 
first nucleus in the A+A system (Cu) to be positive. An even more recent paper from 
STAR reports $v_1(p_T)$ distributions for the same system and centrality that are 
consistent within errors~\cite{star_CuAu}.  

The PHENIX results in Fig.~\ref{phenix_CuAu_pT}~\cite{phenix_CuAu} reveal that the 
higher $p_T$ particles at midrapidity (above 1 or 1.5 GeV/$c$ in $p_T$), and at all 
the studied centralities, have negative $v_1$ and so are preferentially emitted 
with azimuths parallel to the Au fragment bounce-off direction (and antiparallel to 
the Cu fragment bounce-off direction). Whether or not the more abundant particles 
below 1 GeV/$c$ are preferentially emitted with opposite azimuths, as might be 
expected based on momentum conservation, cannot be answered within the 
systematic uncertainty of the measurements~\cite{phenix_CuAu}.          

\begin{figure}
\begin{center}
\includegraphics[scale=0.35]{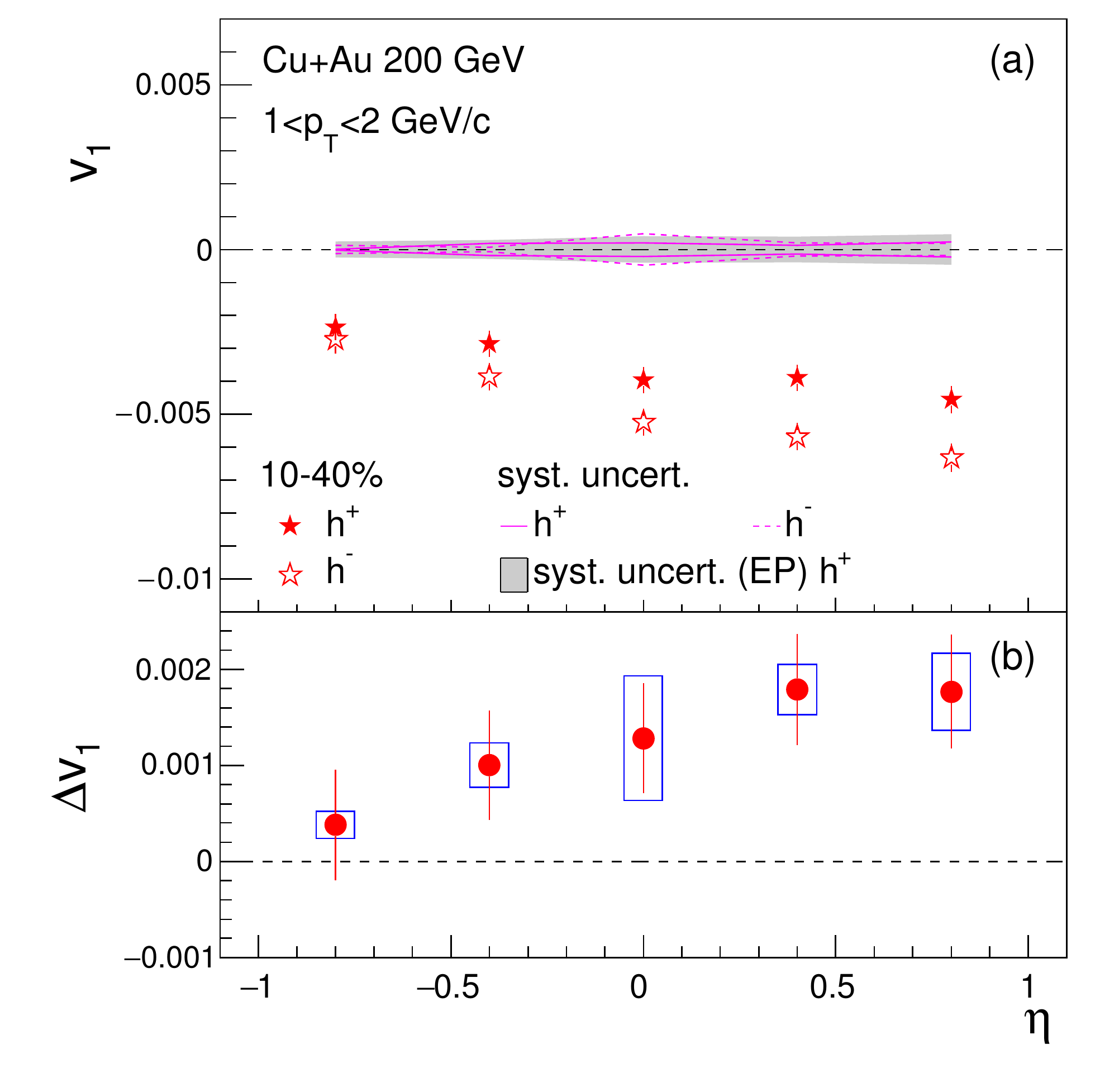}
\caption{(Color online) Directed flow as a function of pseudorapidity, separately 
evaluated for positively and negatively charged particles, at $1 < p_T < 2$ GeV/$c$ 
and centrality 10-40\% in 200 GeV Cu+Au collisions in the STAR detector  
\cite{star_CuAu}. }
\label{star_CuAu_eta}
\end{center}
\end{figure} 

In the STAR collaboration's analysis of charged particle directed flow in Cu+Au 
collisions at $\sqrt{s_{NN}} = 200$ GeV, a particular focus is the $v_1$ difference 
between positive and negative charges. This difference has the potential to be 
sensitive to the strong electric field between the two incident ions, whose 
electric charges differ by $79 - 29 = 50$ units; this field has a lifetime on the 
order of a fraction of a fm/$c$. Fig.~\ref{star_CuAu_eta}~\cite{star_CuAu} presents 
$v_1(\eta)$ at medium $p_T$ ($1 < p_T < 2$ GeV/$c$) and intermediate centrality 
(10-40\%). Like the PHENIX result, this $v_1$ measurement was made relative to the 
event plane from spectator neutrons, dominated by the Au side. It is evident from 
Fig.~\ref{star_CuAu_eta} that both even and odd components are present, and most 
interestingly, there is a significant pattern showing a larger magnitude for 
negative particles. 

The Parton-Hadron String Dynamics (PHSD) model~\cite{phsd_details, phsd_CuAu}, 
when the initial electric field is explicitly modeled, predicts a $v_1$ difference 
signal that is an order of magnitude larger than observed~\cite{star_CuAu}. On the 
other hand, parton distribution functions \cite{hera-pdf} can be used to estimate 
the number of quarks and antiquarks at very early times in relation to the number 
created in the collision; then given certain plausible assumptions, as set out in 
Ref.~\cite{star_CuAu}, it can be inferred that only a small fraction of the total 
quarks created in the collision are produced during the lifetime of the initial 
electric field. In addition to this important insight, the charged-dependent directed 
flow measurements in Cu+Au collisions offer new and valuable quantitative 
information with relevance to the Chiral Magnetic Effect \cite{cme1, cme2} and the 
Chiral Magnetic Wave \cite{cmw1, cmw2}.


\section{Differential measurements of identified particle directed flow }
\label{s-PID}
The charged particle measurements reviewed in Section \ref{s-Charged} are an 
admixture of all emitted particle species. Measurements of directed flow for 
identified particles offer more insights into the underlying physics that 
controls this observable. In this section, we discuss the dependence of $v_1$ 
on $p_T$, $y$ and centrality for several identified particle species.   

\subsection{Dependence of $v_1$ on transverse momentum}

\begin{figure}
\begin{center}
\includegraphics[scale=0.5]{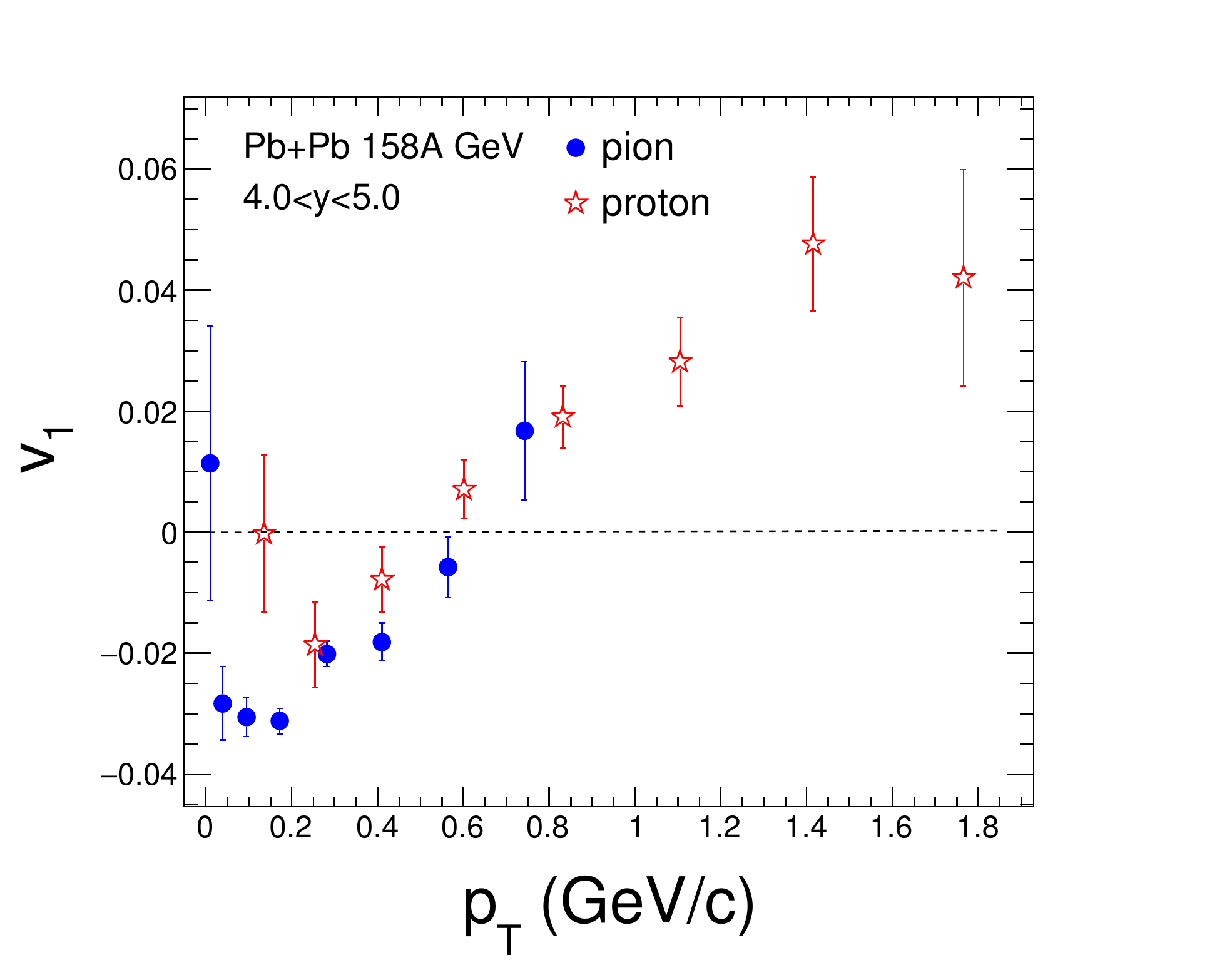}
\caption{(Color online) Pion and proton $v_1(p_T)$ at $4 < y_{\rm lab} < 5$ in 
Pb+Pb collisions at the SPS, at a projectile kinetic energy of 158$A$ GeV 
\cite{na49_prl_80}. }
\label{na49_proton_pion_pT}
\end{center}
\end{figure} 
Measurements of $v_1(p_T)$ for protons, antiprotons and charged pions have been 
reported by the E877 collaboration at the AGS (11$A$ GeV/$c$)~\cite{E877_npa_590, 
E877_prc_56, E877_prc_59, E877_plb_485}. For antiprotons, large negative values 
of $v_1$ are observed for $p_T > 0.1$ GeV/$c$ but with large statistical errors. 
For protons and charged pions, $v_1(p_T)$ in various rapidity gates have been 
published, and these results are also divided into various bins of transverse 
energy, $E_T$, which is a proxy for centrality. The E877 collaboration also 
provides the information needed to convert from their intervals of $E_T$ into 
percent centrality \cite{E877_prc_56}.  

The NA49 collaboration~\cite{na49_prl_80} measured proton and pion $v_1(p_T)$ in 
Pb+Pb collisions at a projectile kinetic energy of 158$A$ GeV, as shown in 
Fig.~\ref{na49_proton_pion_pT}. The rapidity gate for these measurements is 
$4 < y_{\rm lab} < 5$, which corresponds to a forward region (midrapidity is 
$y_{\rm lab} = 2.92$). The NA49 collaboration describes the $v_1(p_T)$ behavior 
as `peculiar', especially for pions.  However, they point out that negative $v_1$ 
at low $p_T$ has been predicted by Voloshin \cite{Voloshin97}, and is explained 
by the interaction of radial and directed flow. Various types of non-flow effect 
were also mentioned as possible contributors to the observed pion behavior 
\cite{na49_prl_80}.

\subsection{Dependence of $v_1$ on rapidity}

\begin{figure}
\begin{center}
\includegraphics[scale=0.7]{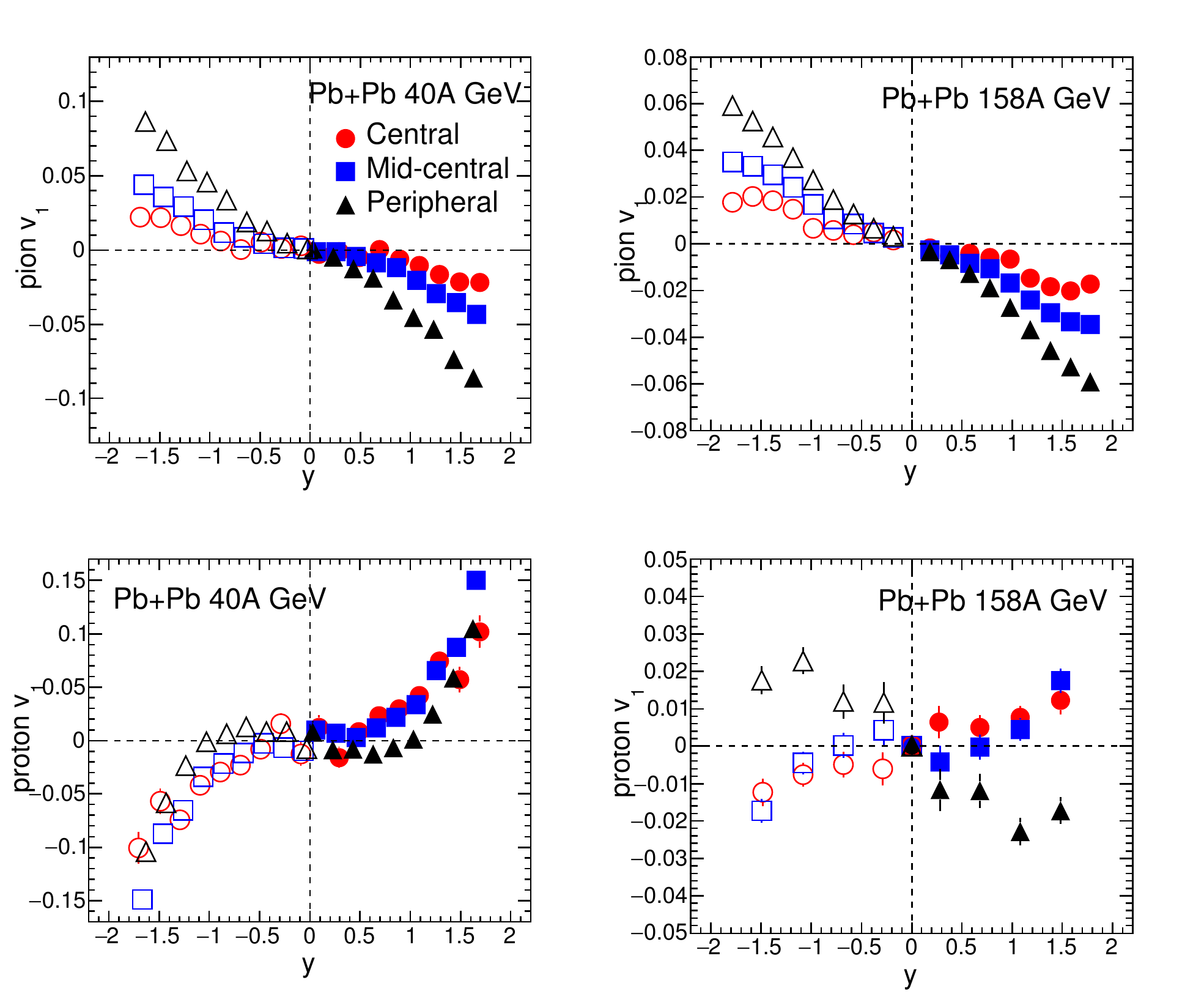}
\caption{(Color online) Pion and proton $v_1(y)$ for three centralities in 40$A$ and 
158$A$ GeV Pb+Pb collisions at the SPS~\cite{na49_prc_68}. The open markers at negative 
rapidities were obtained by reflecting the solid markers at positive rapidities, where 
the detector acceptance was optimum.}
\label{na49_proton_pion}
\end{center}
\end{figure}
\begin{figure}
\begin{center}
\includegraphics[scale=0.4]{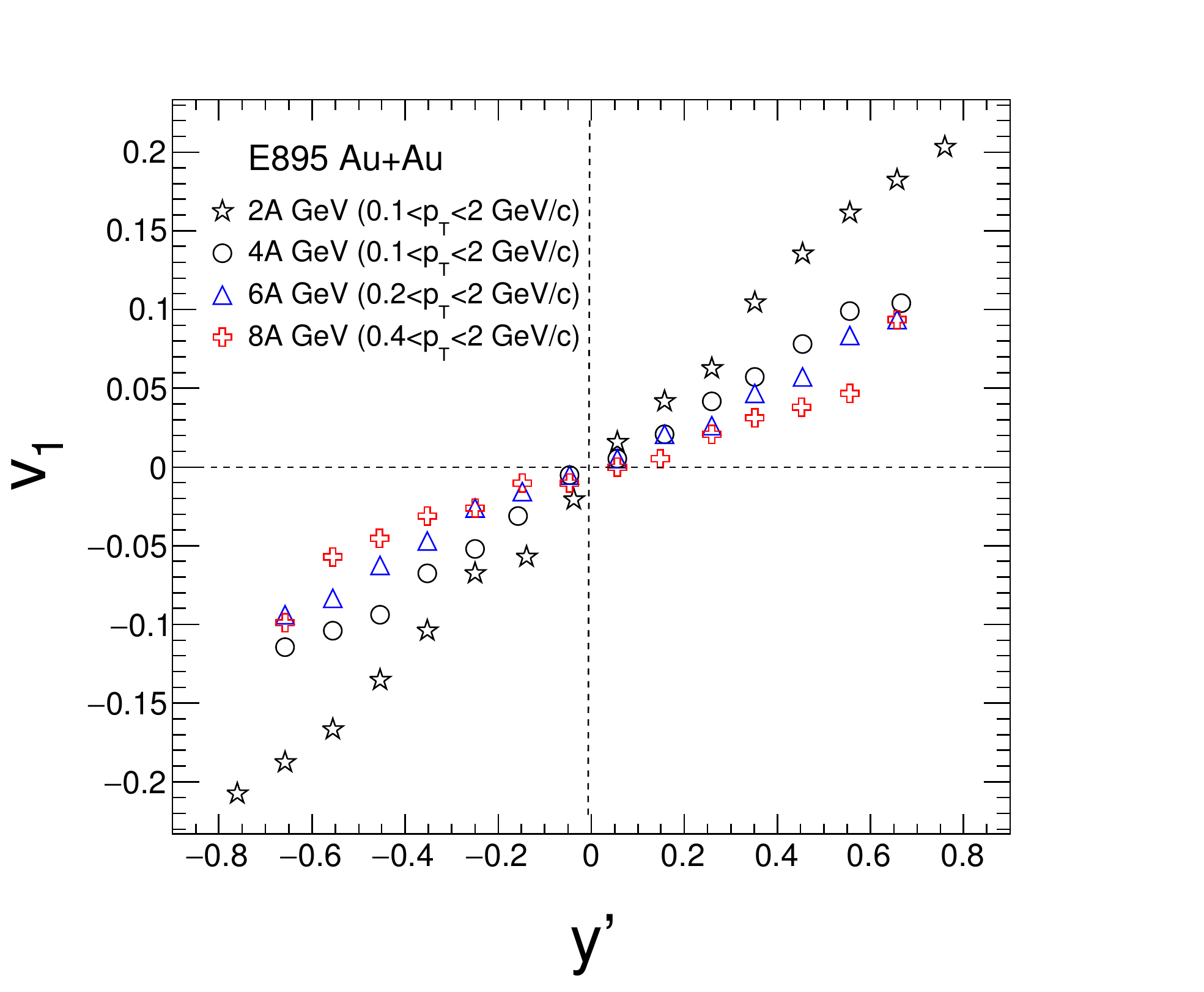}
\includegraphics[scale=0.4]{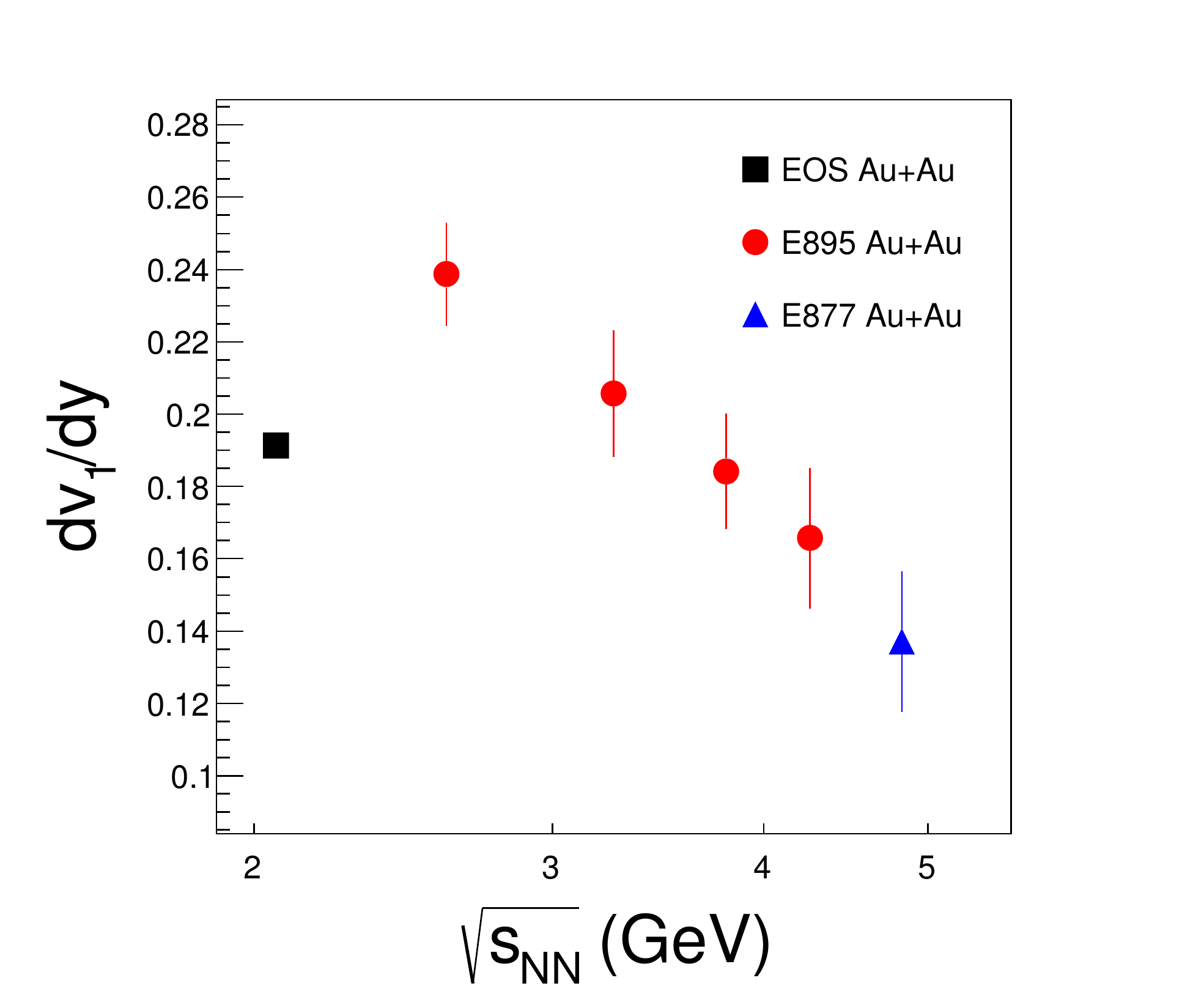}
\caption{(Color online) Left panel: $v_1(y^\prime)$ for protons in 2$A$, 
  4$A$, 6$A$ and 8$A$ GeV Au+Au collisions measured by E895~\cite{E895_prl_84}. 
  Right panel: Beam energy dependence of the slope $dv_1/dy$ (here using 
  un-normalized rapidity) for protons, re-plotted using the more ubiquitous 
  beam energy scale $\sqrt{s_{NN}}$~\cite{E895_prl_84}. A point from the 
  same detector at the Bevalac~\cite{eos_prl_75} and another from E877 at the  
  top AGS energy~\cite{E877_prc_56} are also plotted. }
\label{e895_all}
\end{center}
\end{figure}

 
Various models suggest that the structure of $v_1(y)$ near midrapidity, 
especially the pattern for baryons, is sensitive to the QCD equation of 
state and therefore this signal can be used to investigate QGP production 
and changes of phase \cite{earlytime, h_ritter_review, Norbert_review, 
stocker_npa_750, Heinz2003, Pasi2006}. Fig.~\ref{na49_proton_pion} 
presents proton and pion $v_1(y)$ in central, mid-central and peripheral 
Pb+Pb collisions at projectile kinetic energies of 40$A$ and 158$A$ 
GeV~\cite{na49_prc_68}, as reported by the NA49 collaboration at the CERN SPS. 
The data points at negative rapidity are mirrored from the positive side. The 
$v_1(y)$ for pions is similar in magnitude and shape at both 40$A$ and 158$A$
GeV. The proton $v_1(y)$ measurements suggest that proton anti-flow (also called 
wiggle) \cite{Csernai, Brachmann, snellings_prl_84, Norbert_review, stocker_npa_750}, 
which is not observed at the AGS (see Fig.~\ref{e895_all}), might begin to happen at 
SPS energies. However, based on detailed studies using a variety of methods, 
including the approach introduced by Borghini {\it et al.} \cite{Borghini}, the 
NA49 collaboration reports that the observed pattern of proton $v_1(y)$ at SPS 
energies could be influenced by non-flow effects~\cite{na49_prc_68}.

The EOS-E895 experiment carried out a beam energy scan at the Brookhaven AGS 
during 1996. E895 featured the first Time Projection Chamber with pad readout, and 
reported directed flow in the form of $v_1$, as well as in the form of the older observable 
$\langle p_x \rangle$ \cite{Pawel_Px} (in-plane transverse momentum) for several 
identified particle species: $p$, $\bar{p}$, $\Lambda$, $K_S^0$ and $K^\pm$, in 
Au+Au collisions at projectile kinetic energies 2$A$, 4$A$, 6$A$ and 8$A$ GeV 
\cite{E895_prl_84, E895_prl_85, E895_prl_86, E895_npa_698}. The left panel in 
Fig.~\ref{e895_all} shows $v_1(y^\prime)$ for protons at the four E895 beam 
energies, where $y^\prime$ denotes normalized rapidity such that the target and 
projectile are always at $y^\prime = -1$ and $+1$, respectively. The slope of 
$v_1$ remains positive for all four beam energies.  

The slope of the rapidity dependence of directed flow was extracted by E895 
using a cubic fit $v_1(y^\prime) = F{y^\prime} + Cy^{\prime\;3}$. The right 
panel of Fig.~\ref{e895_all} shows the fitted proton slope $dv_1/dy$ (not 
using normalized rapidity), and unlike in Ref.~\cite{E895_prl_84}, the 
horizontal axis in the right panel of Fig.~\ref{e895_all} uses the 
now-conventional beam energy scale $\sqrt{s_{NN}}$. Two additional points 
for proton $dv_1/dy$ in Au+Au collisions are also plotted here: a measurement 
at the top energy of the Berkeley Bevalac (1.2$A$ GeV)~\cite{eos_prl_75} using 
the same EOS TPC detector as in E895, and an E877 measurement at the top AGS 
energy of 11$A$ GeV \cite{E877_prc_56}. The plotted data show a peak near $2A$ 
GeV beam kinetic energy ($\sqrt{s_{NN}} \sim 2.7$ GeV), and thereafter, a smooth 
decrease with beam energy across the AGS range.  

Three-fluid hydrodynamic calculations~\cite{stocker_npa_750} predict a ``softest
point" in the Equation of State in the AGS energy range, but the E895 beam energy 
scan did not reveal any non-monotonic behavior in the energy dependence. Overall, 
hadronic models with a momentum-dependent mean field~\cite{jam_prc_72} show 
better agreement with data at AGS and SPS energies.

\begin{figure}
\begin{center}
\includegraphics[scale=0.7]{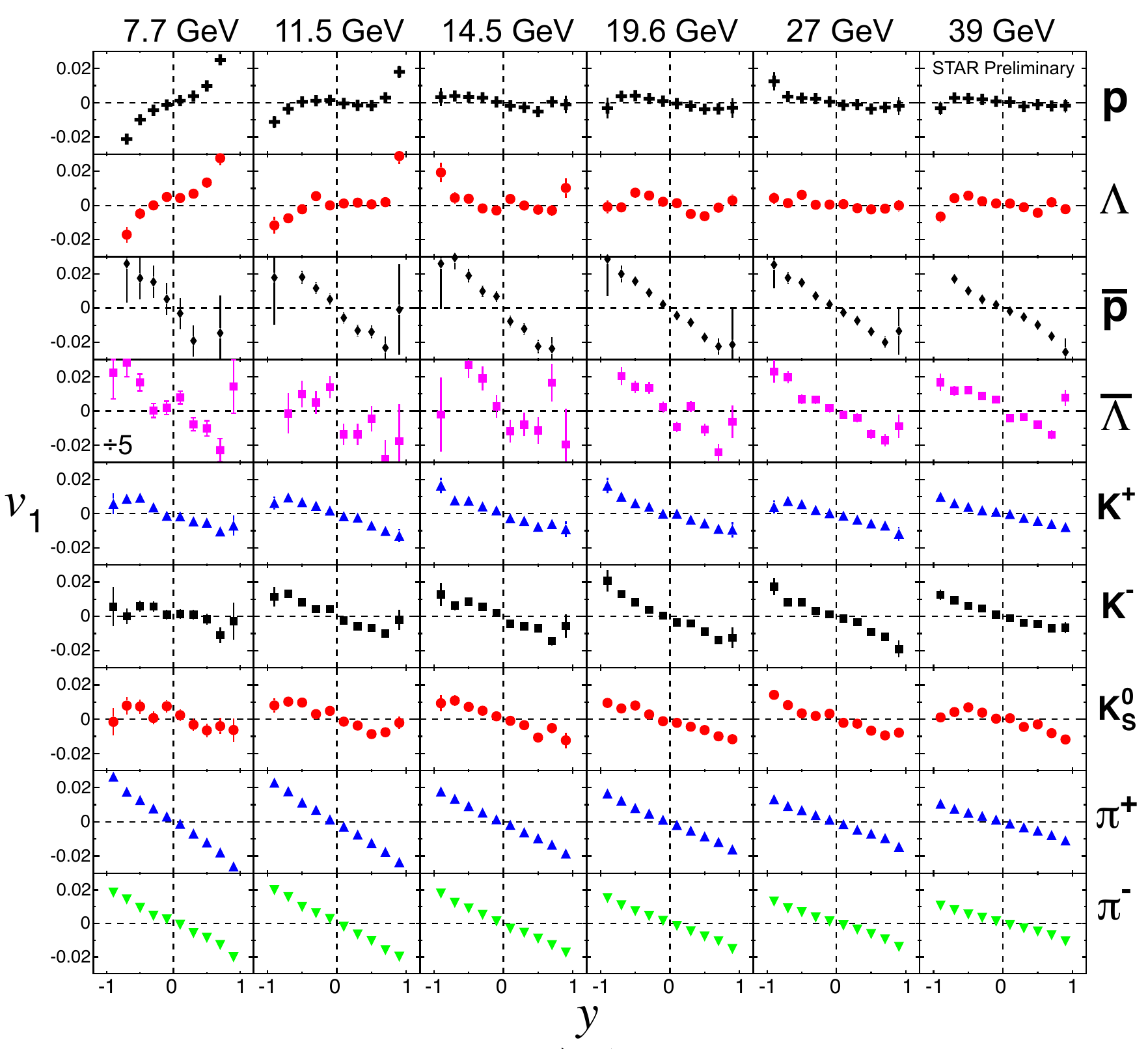}
\caption{(Color online) Directed flow as a function of rapidity for $p$, $\Lambda$, 
$\bar{p}$, $\bar{\Lambda}$, $K^{\pm}$, $K_S^0$, and $\pi^{\pm}$ in 10-40\% centrality 
Au+Au collisions, at $\sqrt{s_{NN}}$ values of 7.7, 11.5, 14.5, 19.6, 27 and 39 GeV 
\cite{star_prl_112, star_qm_2015}. The magnitude of $v_1$ for $\bar{\Lambda}$ at 7.7 
GeV is divided by 5 to fit on the same vertical scale as all the other panels. }
\label{v1vsy_qm15}
\end{center}
\end{figure}

Phase-I of the beam energy scan (BES) \cite{BES1, BES2} at RHIC took data in 
2010, 2011 and 2014, spanning the $\sqrt{s_{NN}}$ range from 200 GeV down to 
7.7 GeV.  Lattice QCD calculations \cite{LQCD1, LQCD2, LQCD3}
imply a smooth crossover from hadronic matter to QGP at beam energies near and 
above the top energy of RHIC, whereas phase diagram features like a first-order 
phase transition and a critical point may become evident in nuclear collisions 
as the beam energy is scanned across RHIC's BES region. The STAR experiment 
initially reported measurements of directed flow for protons, antiprotons and 
charged pions in the energy range of 7.7 to 200 GeV~\cite{star_prl_112}. At the 
2015 Quark Matter meeting, preliminary directed flow data at BES energies for 
nine particle species were presented, along with $v_1(y)$ slopes for protons, 
$\Lambda$ and $\pi^+$ in 9 bins of centrality~\cite{star_qm_2015}. 
Fig.~\ref{v1vsy_qm15} shows $v_1(y)$ at intermediate centrality (10-40\%) for 
$p$, $\Lambda$, $\bar{p}$, $\bar{\Lambda}$, $K^{\pm}$, $K_S^0$, and $\pi^{\pm}$ 
at 7.7 to 39 GeV. 

\begin{figure}
\begin{center}
\includegraphics[scale=0.6]{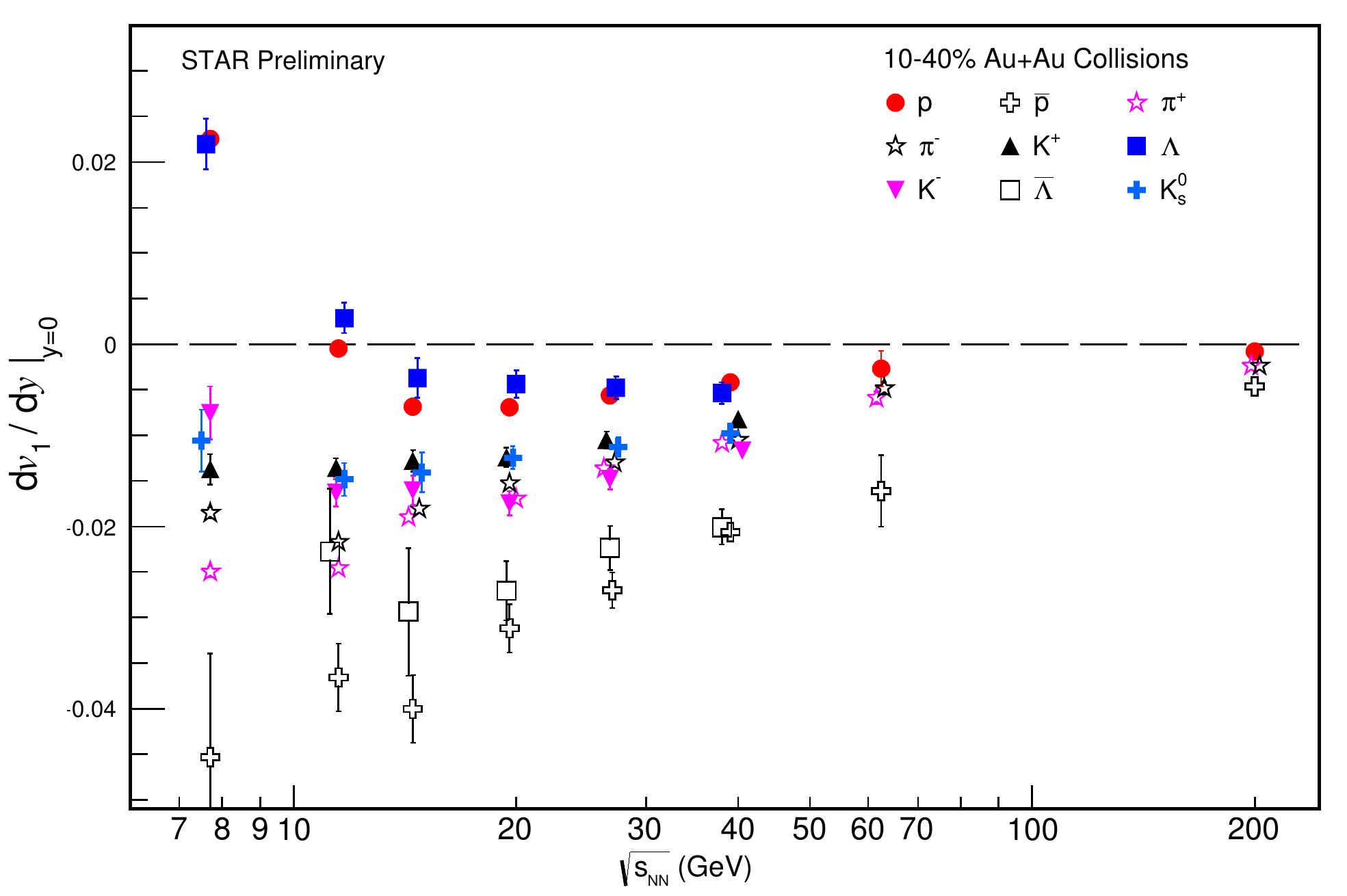}
\caption{(Color online) Beam energy dependence of $v_1(y)$ slope for $p$, $\Lambda$, 
  $\bar{p}$, $\bar{\Lambda}$, $K^{\pm}$, $K_S^0$, and $\pi^{\pm}$ in 10-40\% 
  centrality Au+Au collisions~\cite{star_prl_112, star_qm_2015}.}
\label{dv1dy_qm15}
\end{center}
\end{figure}

The overall strength of directed flow has been characterized by the slope $dv_1/dy$ 
from a linear fit over the range $-0.8 < y < 0.8$~\cite{star_qm_2015}, and the beam 
energy dependence of these slopes for the same nine particle species ($p$, $\Lambda$, 
$\bar{p}$, $\bar{\Lambda}$, $K^{\pm}$, $K_S^0$, and $\pi^{\pm}$) in 10-40\% centrality 
Au+Au collisions are presented in Fig.~\ref{dv1dy_qm15}.  

The most noteworthy feature of the data is a minimum within the $\sqrt{s_{NN}}$ range 
of 10 to 20 GeV in the slope $dv_1/dy|_{y \sim 0}$ for protons at intermediate 
centrality. The same quantity for $\Lambda$ hyperons is consistent with the proton 
result, but the larger statistical errors for $\Lambda$ do not allow an independent 
determination of a possible minimum in the beam energy dependence for this species. 
The proton and $\Lambda$ directed flow slope change from positive to negative close 
to 11.5 GeV, and remain negative at all the remaining energies (up to and including 
200 GeV in the case of protons). The remaining species have negative slope at all 
the studied beam energies. 
At 7.7 GeV, $dv_1/dy$ for $K^-$ is closer to zero than $dv_1/dy$ for $K^+$, which 
supports the inference that $K^+$ and $K^-$ experience nuclear potentials that are 
repulsive and attractive, respectively, under the conditions created at this beam 
energy \cite{Cassing_Crete2014}. 

\subsection{Net-particle directed flow} 

There are two separate contributions to the energy dependence of proton directed 
flow in the vicinity of midrapidity: one part arises from baryon number transported 
from the initial state at beam rapidity towards $y \sim 0$ by the stopping process 
of the collision, while the other part arises from baryon-antibaryon pairs produced 
in the fireball near midrapidity. Clearly, these two contributions have very 
different dependence on beam energy, and disentangling them has a good potential to 
generate new insights. Towards this end, the STAR collaboration has defined 
\cite{star_prl_112} net-proton directed flow according to 
$$[v_{1}(y)]_p = r(y)[v_{1}(y)]_{\bar{p}} + [1-r(y)]\,[v_{1}(y)]_{{\rm net\mbox{-}}p}$$  
where $r(y)$ is the observed rapidity dependence of the antiproton to proton
ratio. Net-kaon $v_1(y)$ is defined analogously, with $K^+$ and $K^-$ substituted 
for $p$ and $\bar p$, respectively \cite{star_qm_2015}.  

\begin{figure}
\begin{center}
\includegraphics[scale=0.4]{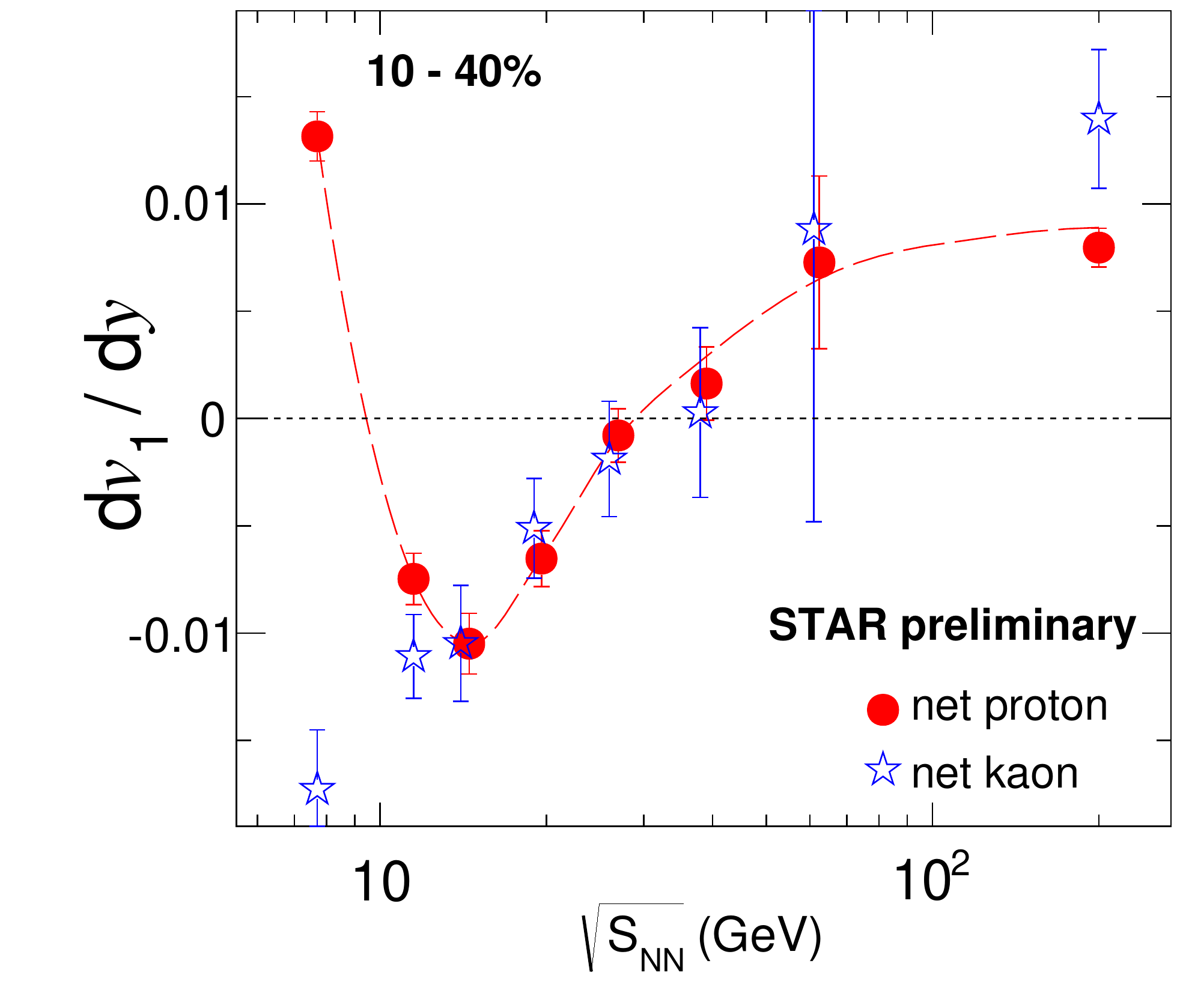}
\caption{(Color online) Beam energy dependence of the slope of $v_1(y)$ for net 
  protons and net kaons in 10-40\% centrality Au+Au collisions, as reported by  
  STAR \cite{star_prl_112, star_qm_2015}.  }
\label{netp_qm15}
\end{center}
\end{figure}

STAR's measurements of net-proton and net-kaon directed flow slope as a function 
of beam energy are reproduced in Fig.~\ref{netp_qm15}~\cite{star_qm_2015}. The 
net-proton directed flow shows a double sign change, and a clear minimum around 
the same beam energy where the proton directed flow has its minimum. This 
coincidence is not surprising, since antibaryon production is low at and below 
the beam energy of the minimum, and therefore net-proton and proton observables 
only begin to deviate from each other at energies above the minimum. 

The observed minimum in directed flow for protons and net-protons resemble the 
predicted ``softest point collapse'' of directed flow \cite{stocker_npa_750, 
jam_attractive}, and these authors are open to an interpretation in terms of a 
first-order phase transition. Other theorists (see Section \ref{s-Models}) point 
out that mechanisms other than a first-order phase transition can cause a drop 
in pressure (a softening of the equation of state) and therefore a definitive 
conclusion requires more research. 

The net-kaon directed flow reproduced in Fig.~\ref{netp_qm15}~\cite{star_qm_2015} 
shows close agreement with the net-proton result near and above 11.5 GeV, but 
deviates very strongly at 7.7 GeV. This deviation is not understood. To date, 
all of the several model comparisons with STAR $v_1$ measurements at BES energies 
have considered only particle $v_1$ as opposed to net-particle $v_1$.  
 
\section{Recent model calculations of directed flow}
\label{s-Models}
Models that explicitly incorporate properties of the Quantum Chromodynamics 
phase diagram and its equation of state (typically hydrodynamic models) suggest 
that the magnitude of directed flow is an excellent indicator of the relative 
pressure during the early, high-density stage of the collision. Therefore, 
directed flow at sub-AGS beam energies can reveal information about hadron gas 
incompressibility, while at the higher energies that are a focus of the present 
review, it can flag the softening, or drop in pressure, that may accompany a 
transition to a different phase, notably Quark Gluon Plasma. For example, there 
may be a spinodal decomposition associated with a first-order phase transition 
\cite{Shukla2001, Bessa2009}, which would cause a large softening effect. However, 
interpretation of flow measurements is not straightforward, and it is known that 
directed flow can also be sensitive to poorly-understood model inputs like 
momentum-dependent potentials \cite{Cassing_Crete2014} in the nuclear medium. 
More theoretical work is needed to elucidate the quantitative connection between 
softening signatures, like the beam energy dependence directed flow, and QCD 
phase changes.  

\begin{figure}
\begin{center}
\includegraphics[scale=0.6]{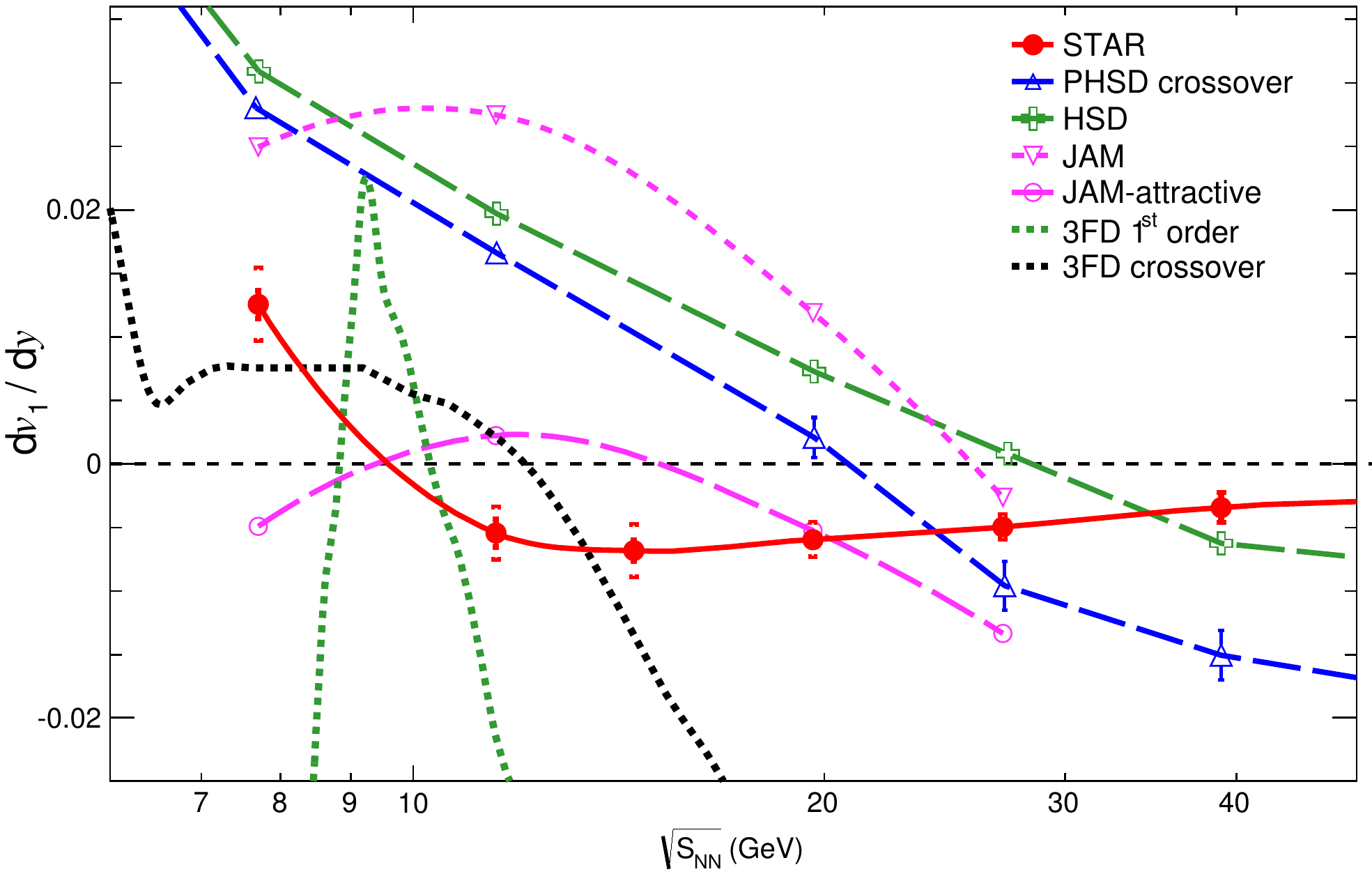}
\caption{(Color online) Beam energy dependence of directed flow slope for protons 
   in 10-40\% centrality Au+Au from the STAR experiment, compared with recent 
   available model calculations \cite{jam_attractive, phsd, Ivanov2015}. All the  
   experimental data are from Ref.~\cite{star_prl_112} except for one energy point, 
   $\sqrt{s_{NN}} = 14.5$ GeV \cite{star_qm_2015}, which should be considered a 
   preliminary measurement. The Frankfurt hybrid model \cite{hybrid_frankfurt} as 
   well as a pure hydro calculation with particle freeze-out at constant energy density 
   \cite{hybrid_frankfurt} both lie above the data and are off-scale at all BES energies.
    }
\label{model_comparison1}
\end{center}
\end{figure} 

During the period since publication of the STAR BES directed flow results in 2014, 
there have been several theoretical papers \cite{hybrid_frankfurt, phsd, Ivanov2015, 
Cassing_Crete2014, jam_attractive} aimed towards interpretation of these 
measurements. The Frankfurt hybrid model \cite{hybrid_details} used for the 
data comparison by Steinheimer {\it et al.}~\cite{hybrid_frankfurt} is based on a 
Boltzmann transport approach similar to UrQMD for the initial and late stages of the 
collision process, while a hydrodynamic evolution is employed for the intermediate 
hot and dense stage. The equation of state for the hydro stage includes crossover 
and first-order phase transition options. The data comparison by Konchakovski {\it 
et al.}~\cite{phsd} uses the Parton-Hadron String Dynamics (PHSD) model 
\cite{phsd_details} of the Giessen group, a microscopic approach with a crossover 
equation of state having properties similar to the crossover of lattice QCD 
\cite{LQCD1, LQCD2, LQCD3}. The PHSD code also has a mode named Hadron String 
Dynamics (HSD), which features purely hadronic physics throughout the collision 
evolution, and which yields directed flow predictions in close agreement 
with those \cite{star_prl_112} of the UrQMD model.  The data comparison by 
Ivanov and Soldatov \cite{Ivanov2015} uses a relativistic 3-fluid hydrodynamic 
model (3FD) \cite{3FD_details} with equations of state that include a crossover 
option and a first-order phase transition option.  The most recent comparison to 
the STAR BES $v_1$ data, by Nara {\it et al.} \cite{jam_attractive}, uses the Jet 
AA Microscopic (JAM) model \cite{JAM_details}. JAM is a purely hadronic 
Boltzmann transport code, but the authors of Ref.~\cite{jam_attractive} 
introduce an option to switch from the normal stochastic binary scattering style 
to a modified style where the elementary 2-body scatterings are always oriented
like attractive orbits \cite{Kahana1995, Kahana1997}. They argue that the 
switch-over from random to attractive binary orbits mimics the softening effect of 
a first-order phase transition.  

Fig.~\ref{model_comparison1} focuses on the most promising directed flow 
measurement from the RHIC Beam Energy Scan, namely the $dv_1/dy|_{y \sim 0}$ 
for protons at 10-40\% centrality, and summarizes recent model comparisons 
\cite{hybrid_frankfurt, phsd, Ivanov2015, jam_attractive} with these data. 
These authors are largely in agreement that the data disfavor models with purely 
hadronic physics. However, some conclude that a crossover deconfinement transition 
is favored \cite{phsd, Ivanov2015}, while others conclude that a first-order phase 
transition is still a possible explanation~\cite{jam_attractive}. Note that the 
argument of Nara {\it et al.}~\cite{jam_attractive} is that a more sophisticated 
implementation of a first-order phase transition would transition from the `JAM' 
curve at low BES energies to the `JAM-attractive' curve at higher BES energies.  

\begin{figure}
\begin{center}
\includegraphics[scale=0.6]{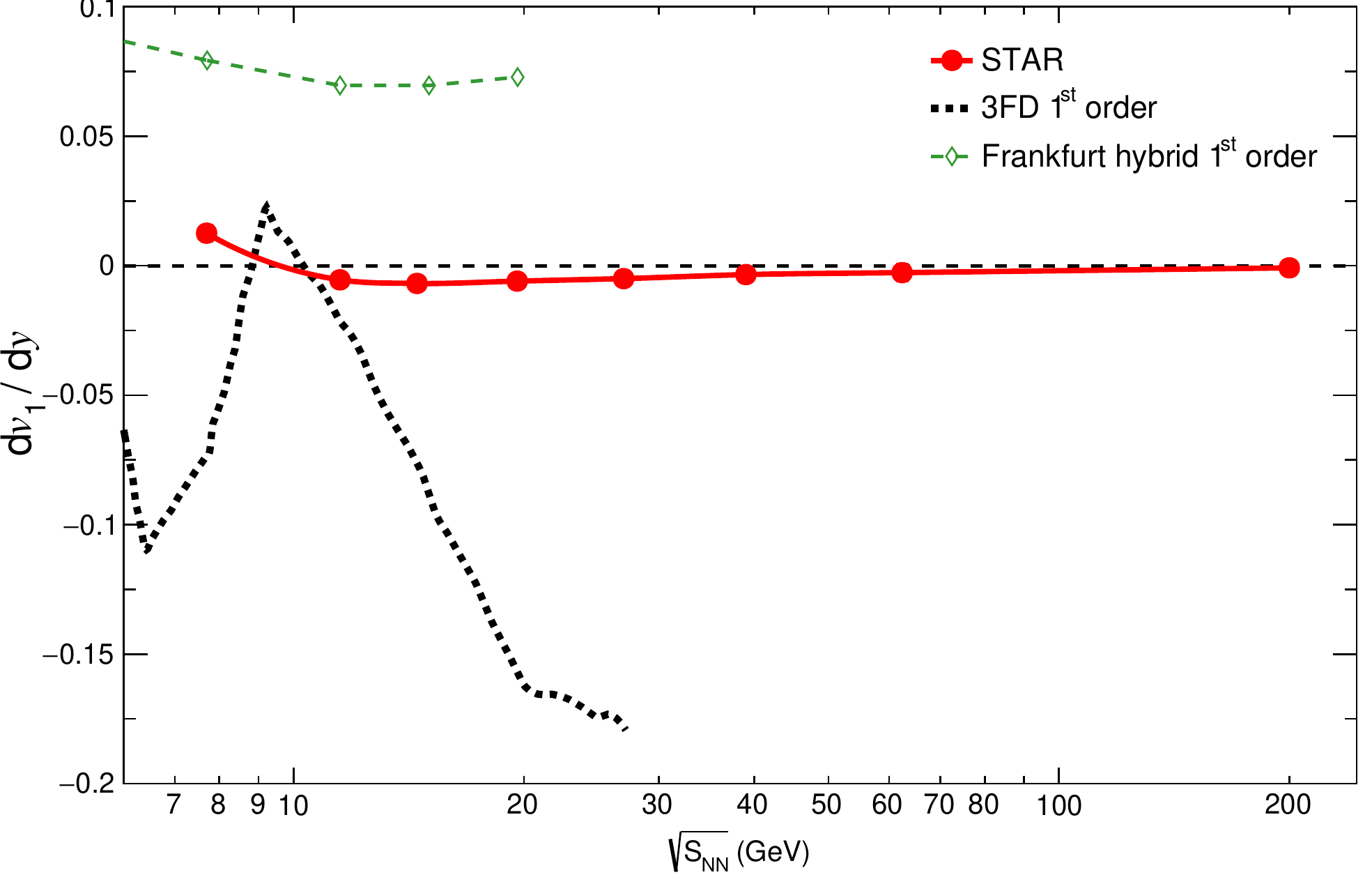}
\caption{(Color online) Beam energy dependence of directed flow slope for protons 
   in 10-40\% centrality Au+Au from the STAR experiment, compared with recent 
   hybrid \cite{hybrid_frankfurt} and 3FD~\cite{Ivanov2015} model calculations. All the  
   experimental data are from Ref.~\cite{star_prl_112} except for one energy point, 
   $\sqrt{s_{NN}} = 14.5$ GeV \cite{star_qm_2015}, which should be considered a 
   preliminary measurement.}
\label{model_comparison2}
\end{center}
\end{figure}

Overall, Fig.~\ref{model_comparison1} underlines the fact that no option in any of 
the model calculations to date reproduces, even qualitatively, the most striking 
feature of the data, namely, the minimum in proton directed flow in the region of 
$\sqrt{s_{NN}} \sim 10 - 20$ GeV.  It is also noteworthy that the $v_1$ 
difference between nominally equivalent equation of state implementations 
in different models is very large. For example,  
the differences in $dv_1/dy$ between the 1st-order phase transition in the hybrid 
model \cite{hybrid_frankfurt} and a similar nominal quantity for the 3FD model 
\cite{Ivanov2015} (see Fig.~\ref{model_comparison2}) is currently more than an 
order of magnitude larger than the experimental measurement being interpreted, 
and is larger still than the error on the measured data.

\section{Summary and outlook}   
\label{s-End}
In this review, we discuss heavy ion directed flow results for charged particles and 
for identified particle types, covering beam energies from the Brookhaven AGS to the 
CERN LHC.  Charged particle directed flow measurements have been published as a 
function of transverse momentum, pseudorapidity, and collision centrality, while 
Cu+Cu and Au+Au have also been compared. The charged particle directed flow 
magnitude at the LHC is a factor of three smaller than that at top RHIC energy. The 
observations from RHIC suggest that the charged particle directed flow is independent 
of system size, but depends on the incident beam energy. Limiting fragmentation 
scaling is observed for $v_1$ at RHIC energies, but entropy-driven multiplicity 
scaling in terms of $dN_{\rm ch}/d\eta$ is not seen at RHIC. In mass-asymmetric 
collisions, specifically Cu+Au, recent directed flow measurements at $\sqrt{s_{NN}} 
= 200$ GeV have opened a new window into quark and antiquark formation at the very 
earliest times of the collision evolution ($t \leq 0.25$ fm/$c$) and could clarify 
theoretical and experimental questions related to the Chiral Magnetic Effect and the 
Chiral Magnetic Wave. 

Measurements of $v_1$ for identified species offer deeper insights into the 
development of hydrodynamic flow. Opposite $v_1$ for pions and protons at AGS/SPS 
and at lower RHIC energies suggests an important role for nuclear shadowing. Signals 
of anti-flow of neutral kaons in AGS/E895 together with kaon measurements in the 
RHIC Beam Energy Scan region point to kaon-nucleon potential effects. The single 
sign-change in proton $v_1$ slope and a double sign-change in net-proton $v_1$ 
slope, with a clear minimum around $\sqrt{s_{NN}} \sim $ 11.5 - 19.6 GeV shows a 
qualitative resemblance to a hydrodynamic model prediction called ``softest point 
collapse of flow''. This original prediction assumed a first-order phase transition,   
but a crossover from hadron gas to a deconfined phase can also cause a softening (a 
drop in pressure). None of the current state-of-the-art models can explain the main 
features of the STAR directed flow measurements, and different models with 
nominally similar equations of state diverge from each other very widely over the 
BES range. 

Looking ahead to likely developments during the period 2017-2018 in the area of 
directed flow at $\sqrt{s_{NN}}$ of a few GeV and above, we can expect new BES 
Phase-I results for the $\phi$ meson, as well as final publication of current 
preliminary RHIC Beam Energy Scan Phase-I results, like those in Ref.~\cite{
star_qm_2015}. We can also expect parallel theoretical work on related physics and 
interpretation of the newest data. The preliminary results for new particle species 
like $\Lambda$s and charged and neutral kaons, as well as BES $v_1$ for protons, 
$\Lambda$s and pions in narrow bins of centrality (nine bins spanning 0-80\% 
centrality) amount to very stringent constraints on the next round of theoretical 
interpretation in terms of the QCD phase diagram and equation of state.  

More comprehensive measurements of the many phenomenological aspects of directed 
flow outlined in this review will be possible beginning in the year 2019, as a 
consequence of the much increased statistics of Phase-II of the RHIC Beam Energy 
Scan (to take data in 2019 and 2020), in conjunction with the anticipated upgrades to 
the performance of the STAR detector \cite{BES-II}. Thereafter, new facilities like FAIR 
\cite{FAIR} in Germany, NICA \cite{NICA} in Russia, and J-PARC-HI~\cite{J-PARC-HI, 
J-PARC-HI-wp} in Japan, will begin coming online. These new dedicated facilities will 
further strengthen worldwide research on the QCD phase diagram at high baryon 
chemical potential. 

~\\

\noindent{\bf Conflict of Interests}\\
The authors declare that there is no conflict of interests regarding
the publication of this paper.
\\

\noindent{\bf Acknowledgments}\\
This work was supported in part by the Office of Nuclear Physics within the US DOE 
Office of Science, under grant DE-FG02-89ER40531. We are grateful to the following 
colleagues for insightful advice and discussions: J. Auvinen, H. Caines, D. Cebra, 
V. Dexheimer, M. Lisa, W. Llope, G. Odyniec, Y. Pandit, H. Petersen, A. Poskanzer, 
H. G. Ritter, J. Steinheimer, H. St\"ocker, A. H. Tang, S. Voloshin, G. Wang, N. Xu, 
and Z. Xu.

\end{document}